\documentclass[twocolumn,twocolappendix,,trackchanges]{aastex701}
\usepackage{multirow}
\usepackage{float}
\usepackage{layouts}
\usepackage{comment}
\usepackage{placeins}
\usepackage{xspace}

\newcommand{\posydon}{\texttt{POSYDON}\xspace}

\begin{document}

\title{Stellar Population Spectra Incorporating Detailed Binary Evolution using POSYDON}
\correspondingauthor{Eirini Kasdagli}
\email{kasdaglie@ufl.edu}

\author[0009-0009-1888-8785]{Eirini Kasdagli}
\email{kasdaglie@ufl.edu}
\affiliation{Department of Physics, University of Florida, 2001 Museum Rd, Gainesville, FL 32611, USA}

\author[0000-0001-5261-3923]{Jeff J. Andrews}
\email{jeffrey.andrews@ufl.edu}
\affiliation{Department of Physics, University of Florida, 2001 Museum Rd, Gainesville, FL 32611, USA}
\affiliation{Institute for Fundamental Theory, 2001 Museum Rd, Gainesville, FL 32611, USA}

\author[0000-0003-2192-3296]{Bret Lehmer}
\email{lehmer@uark.edu}
\affiliation{Department of Physics, University of Arkansas, 226 Physics Building, 825 West Dickson Street, Fayetteville, AR 72701, USA}

\author[0000-0002-2522-8605]{Richard H. D. Townsend}
\email{townsend@astro.wisc.edu}
\affiliation{Department of Astronomy, 2535 Sterling Hall, University of Wisconsin-Madison, Madison, WI 53706, USA}

\author[0000-0002-7464-498X]{Manos\,Zapartas}\email{ezapartas@gmail.com}
\affiliation{ Physics Department, National and Kapodistrian University of Athens, 15784 Athens, Greece}
\affiliation{Institute of Astronomy, Foundation for Research and Technology -- Hellas, Voutes, 71110 Heraklion, Greece}

\author[0000-0001-8952-676X]{Andreas Zezas}
\email{azezas@physics.uoc.gr}
\affiliation{Institute of Astronomy, Foundation for Research and Technology -- Hellas, Voutes, 71110 Heraklion, Greece}
\affiliation{Physics Department \& Institute of Theoretical \& Computational Physics, University of Crete, 71003 Heraklion, Crete, Greece}

\author[0000-0002-6960-6911]{Ylva\,Götberg}
\email{ylva.gotberg@ista.ac.at}
\affiliation{Institute of Science and Technology Austria (ISTA), Am Campus 1, 3400 Klosterneuburg, Austria}

\author[0000-0002-6842-3021]{Max\,Briel}\email{email}
\affiliation{Département d’Astronomie, Université de Genève, Chemin Pegasi 51, CH-1290 Versoix, Switzerland}
\affiliation{Gravitational Wave Science Center (GWSC), Université de Genève, CH1211 Geneva, Switzerland}

\author[0000-0003-1474-1523]{Tassos\,Fragos}\email{email}
\affiliation{Département d’Astronomie, Université de Genève, Chemin Pegasi 51, CH-1290 Versoix, Switzerland}
\affiliation{Gravitational Wave Science Center (GWSC), Université de Genève, CH1211 Geneva, Switzerland}

\author[0000-0001-6692-6410]{Seth\,Gossage}\email{email}
\thanks{Co-author as core developer of \posydon.}
\affiliation{Center for Interdisciplinary Exploration and Research in Astrophysics (CIERA), Northwestern University, 1800 Sherman Ave, Evanston, IL 60201, USA}
\affiliation{NSF-Simons AI Institute for the Sky (SkAI),172 E. Chestnut St., Chicago, IL 60611, USA}

\author[0000-0003-1749-6295]{Philipp\,M.\,Srivastava}\email{email}
\thanks{Co-author as core developer of \posydon.}
\affiliation{Center for Interdisciplinary Exploration and Research in Astrophysics (CIERA), Northwestern University, 1800 Sherman Ave, Evanston, IL 60201, USA}
\affiliation{NSF-Simons AI Institute for the Sky (SkAI),172 E. Chestnut St., Chicago, IL 60611, USA}
\affiliation{Electrical and Computer Engineering, Northwestern University, 2145 Sheridan Road, Evanston, IL 60208, USA}

\author[0000-0003-0420-2067]{Elizabeth\,Teng}\email{email}
\thanks{Co-author as core developer of \posydon.}
\affiliation{Center for Interdisciplinary Exploration and Research in Astrophysics (CIERA), Northwestern University, 1800 Sherman Ave, Evanston, IL 60201, USA}
\affiliation{NSF-Simons AI Institute for the Sky (SkAI),172 E. Chestnut St., Chicago, IL 60611, USA}
\affiliation{Department of Physics and Astronomy, Northwestern University, 2145 Sheridan Road, Evanston, IL 60208, USA}



\begin{abstract}
The accuracy of stellar population properties inferred through spectral energy distribution fitting hinges on the reliability of the underlying spectral models. Binary interactions are fundamental for massive star evolution, and ignoring their spectral contribution can lead to incorrect results. We use the POSYDON binary population synthesis code to generate spectral models of stellar populations that include binaries at solar metallicity. Our framework incorporates a collection of spectral libraries that is designed to address key outcomes of binary stellar evolution like wind stripped Wolf-Rayet stars, binary stripped helium stars, and a treatment for stellar mergers. Our starburst spectral models confirm previous results showing that the inclusion of binary interactions has a significant effect on the UV and ionizing regime of the integrated spectrum. In particular we find that Wolf-Rayet stars and other massive main sequence stars dominate the production of ionizing radiation at earlier times, but after $\simeq$16 Myr binary stripped stars produced through mass transfer begin to dominate. Furthermore, we show that the production of ionizing $\text{He}^+$ photons is especially sensitive to the underlying population of binary stripped stars. While our results currently focus on high-mass stars ($\ge4~M_{\odot}$) at Solar metallicity, they provide the framework for binary spectral synthesis across a range of metallicities and masses and lay the foundation for calculations of the emergent emission-line spectra in the UV, optical, and IR regimes.
We make the spectral models from this work publicly available for use in a format that can be integrated into fitting codes. 
\end{abstract}

\keywords{}


\section{Introduction} \label{sec:intro}

    Stellar populations beyond the Local Universe are typically observed not as collections of individual stars but through their integrated light. To further our understanding of stellar and galaxy evolution, as well as the properties of stellar populations, we must extract and disentangle the wealth of information imprinted in their spectra. The most common technique used for this purpose is to fit the observations with predictions from theoretical stellar population synthesis (SPS) models.

    %
    In its simplest form, SPS combines the initial mass function (IMF), stellar evolutionary tracks (e.g. isochrones), and stellar atmospheres to construct the integrated light of stellar populations \citep[see][for a review]{2013ARA&A..51..393C}. 
     First efforts that implemented this framework to create SPS models include \cite{1968ApJ...151..547T}, and as stellar evolution progressed, it led the way for modern widely used spectral synthesis codes such as {\tt PEGASE} \citep{1997A&A...326..950F,1999astro.ph.12179F,2004A&A...425..881L}, {\tt STARBURST99} \citep{1999ApJS..123....3L,2010ApJS..189..309L,2014ApJS..212...14L,2025ApJS..280....5H}, and {\tt GALAXEV} \citep{2003MNRAS.344.1000B}. These SPS models are then combined with a star formation history (SFH) and nebular and dust components, to create SEDs of composite populations such as galaxies. Tools such as {\tt FSPS} \citep{2009ApJ...699..486C} provide a flexible framework for incorporating these ingredients into SED models. The resulting SEDs are then used to infer galaxy properties by comparing model predictions with observations. Inference tools, such as \texttt{PROSPECTOR} \citep{2017ApJ...837..170L,2021ApJS..254...22J}, \texttt{BAGPIPES} \citep{2018MNRAS.480.4379C}, \texttt{LIGHTNING} \citep{2024ApJ...977..189L}, \texttt{BEAGLE} \citep{2016MNRAS.462.1415C}, and \texttt{CIGALE} \citep{2019A&A...622A.103B}, are then used on photometric or spectroscopic data to extract properties like the total stellar mass, ages, SFH, and dust properties. 
    
    In recent years, continuous advancements made in stellar evolution and atmospheric radiative transfer modeling have been progressively incorporated into SPS codes. Many of these improvements have focused on previously neglected phases of stellar evolution and have highlighted how sensitive the inferred properties of stellar populations are to the underlying assumptions of SPS models. In particular, a more careful treatment of thermally pulsing asymptotic giant branch (TP-AGB) stars led to systematic differences in the predicted stellar masses of intermediate-age populations \citep{2007ASPC..374..303B,2009ApJ...699..486C}.
    Advances in massive star models, including updated wind prescriptions and improved atmospheres for Wolf-Rayet (WR) stars \citep{2015MNRAS.452.1068C}, have been incorporated into efforts to improve ultraviolet (UV) output and emission line predictions of SPS \citep{2016MNRAS.462.1757G}. 
    Additionally, modeling the effects of stellar rotation and extending the IMF to higher masses and lower metallicities has led to an increase in the production of ionizing radiation predicted by SPS models \citep{2012ApJ...751...67L,2014ApJS..212...14L,2015ApJ...800...97T,2017ApJ...838..159C,2025ApJS..280....5H}.

    Despite these improvements, one of the most important aspects of massive star evolution that the majority of SPS models fail to account for is binary evolution. Yet, binary stars are ubiquitous. Recent surveys of resolved stellar populations have shown that the majority of massive stars are found gravitationally bound to at least one other star \citep{2012Sci...337..444S, 2017ApJS..230...15M} in close orbits \citep{2014ApJS..213...34K} where they can eventually interact with their companion. Their impact on population spectra is particularly profound since binary interactions alter the evolution of the individual stars, producing stellar outcomes inaccessible through single star evolution. In particular, stellar outcomes of binary evolution, such as stripped-envelope stars \citep{1999ARA&A..37..603O,2019A&A...629A.134G}, rapid rotators \citep{2017PASA...34...58E, 2023MNRAS.518..860G}, and stellar mergers \citep{2019Natur.574..211S}, can produce copious amounts of UV and ionizing photons \citep{2018MNRAS.479...75S,2024MNRAS.527.9480L}.

    UV and ionizing photon production are key for interpreting galaxy properties. For example, observations of the UV continuum and nebular emission lines in galaxy SEDs, in tandem with predictions from SPS models, have been widely used to estimate star formation rates in galaxies \citep{1998ARA&A..36..189K,2007ApJS..173..267S} and to constrain the cosmic star formation history \citep{2014ARA&A..52..415M,2019MNRAS.490.5359W}. Furthermore, ionizing radiation produced from SPS models has been used to compute the emerging emission-line spectra of different stellar populations, subsequently providing emission-line diagnostics that can distinguish between star formation and AGN activity \citep{2019ARA&A..57..511K}. The UV is also critical for modeling the full SED of galaxies under the ``energy balance" principle, in which the mid- and far-IR emission results from the reprocessed UV/optical radiation by interstellar dust \citep{2019A&A...622A.103B,2013ARA&A..51..393C}. Under this assumption the luminosity of the IR emission is equal to the absorbed UV/optical radiation after accounting for any escaping radiation, while the shape of the IR SED is driven by dust properties. Finally, by relating SPS models either with their ionizing photon production efficiency $\xi_{ion} $ \citep{2016ApJ...831..176B,2023MNRAS.524.2312E} or to the escape fraction of ionizing photons $f_{esc}$ \citep{2019ApJ...878...87F}, UV observations are used to gain insights into the sources that re-ionize the universe.

    Efforts to incorporate binary interactions into spectral synthesis codes have been made by \texttt{BPASS} \citep{2017PASA...34...58E,2018MNRAS.479...75S}, the {\tt GALSEVN} models \citep{2024MNRAS.527.9480L} that were produced using the \texttt{SEVN} \citep{2023MNRAS.524..426I} binary population synthesis (BPS) code, the Yunnan simulations \citep{2015MNRAS.447L..21Z,2012MNRAS.421..743Z}, and the Brussels group \citep{2003A&A...400..429B}. These codes tend to find that binaries produce harder ionizing radiation and stronger emission lines \citep{2016MNRAS.456..485S,2018MNRAS.477..904X} compared with single star models. Specifically, SPS models that include binaries are required to reproduce nebular emission in some star-forming high-redshift galaxies \citep{2016ApJ...826..159S}, or the observed line ratios of He II $\lambda4686/$ $H \beta$  in young galaxies \citep{2024MNRAS.527.9480L}. Moreover, the increased ionizing output of binaries affects the timing of re-ionization \citep{2018MNRAS.479..994R}, with stripped stars in particular providing delayed ionizing radiation (10--200 Myr) that increases the escape fraction of ionizing photons from galaxies \citep{2016MNRAS.459.3614M,2020A&A...634A.134G,2020ApJ...901...72S,2021ApJ...918....5B}. 

    However, incorporating all the outcomes from binary evolution into SPS is both challenging and computationally expensive. Codes mitigate the limitations introduced by the computational costs of solving the stellar structure equations of the two stars along with their orbit by adopting certain approximations and assumptions. The Yunnan simulations and {\tt GALSEVN} rely upon fitting formulae provided by \citet{2000MNRAS.315..543H,2002MNRAS.329..897H} or input tables from single star models \citep{2017MNRAS.470.4739S}. {\tt BPASS} takes a different approach entirely, evolving the most massive star's structure simultaneously with the binary's orbit using a custom version of the {\tt STARS} stellar evolution code \citep{1971MNRAS.151..351E, 1995MNRAS.274..964P, 2004MNRAS.353...87E}, while the secondary star's evolution follows from \citet{2000MNRAS.315..543H}. These authors were among the first to demonstrate the importance of detailed binary interactions \citep{2009MNRAS.400.1019E,2012MNRAS.419..479E} that have led to the widespread adoption of their binary SPS models for SED fitting. 

    In this work we expand upon the approach from {\tt BPASS} using the binary population synthesis code {\tt POSYDON} \citep{2023ApJS..264...45F,2025ApJS..281....3A} to generate a library of SPS models at solar metallicity. Based on {\tt MESA} binary and single star grids \citep{2011ApJS..192....3P,2013ApJS..208....4P,2015ApJS..220...15P, 2018ApJS..234...34P, 2019ApJS..243...10P, 2023ApJS..265...15J}, {\tt POSYDON} follows the evolution of both the donor and the accretor along with the orbit, while also modeling the effects of rotation, binary interactions, and
    angular momentum transport on the stellar structure in a physically motivated way. We process the resulting stellar populations with a blend of spectral atmosphere model libraries to produce a set of general-use stellar population spectra across a range of ages.
    
    This paper is organized as follows. In Section \ref{Section 2} we outline our stellar populations and the spectral synthesis framework we use to compute their spectra. In Section \ref{sec:binary_spectral_models} we present our {\tt POSYDON} binary population synthesis models and their comparison with {\tt POSYDON} single star models. In Section \ref{sec:Model_comparison} we show how our models compare with binary and single star models from the literature. We subsequently describe the current limitations present in our models in Section \ref{sec:limitations} and consider future work to address them. Finally, in Section \ref{sec:summary} we conclude by summarizing our findings.

\label{sec:Intro}

\section{The spectral synthesis method}
\label{Section 2}

Our solar metallicity spectral models combine detailed binary population synthesis results from the {\tt POSYDON} code with an array of stellar spectral library grids covering a wide range of stellar properties and evolutionary phases. To produce our population spectra, we first generate large ($\mathcal{O}(10^6)$ per age bin) synthetic populations of single and binary stars at ZAMS following our best understanding of binary star distributions. We then evolve them using {\tt POSYDON} to a pre-determined age to mimic stellar populations of different ages. This procedure necessarily requires that a new population of stars is generated for every age. For each star in each binary system, {\tt POSYDON} provides a $T_{\rm eff}$, log $g$, and surface abundances from which we calculate a spectrum using an array of spectral libraries taken from the literature along with spectral interpolation. As this procedure has some important differences from the production of other population spectral models in the literature, we describe its different components below in detail.
In Section~\ref{sec:binary_population_models} we give an overview of some aspects of the {\tt POSYDON} code and we detail the specific choices made to produce our single star and binary star populations from which our spectral results stem. Then in Section~\ref{sec:spectral_libraries} we describe the spectral synthesis framework as well as the spectral libraries that were used to compute the spectra of binary populations. 

\subsection{Modeling Binary Populations with {\tt POSYDON}}
\label{sec:binary_population_models}
\setcounter{footnote}{0}

Stellar and binary population models comprise the core of spectral synthesis, providing the essential evolutionary prescriptions for simulating large, unresolved stellar populations. To produce our stellar populations, we opt to use {\tt POSYDON}, an open-source detailed binary population synthesis code that generates synthetic binary star populations using \texttt{MESA} grids and on-the-fly calculations. To generate synthetic populations, {\tt POSYDON} starts with an initial distribution of stellar binaries, evolving them through multiple evolutionary phases by interpolating the outcomes of these grids. 
In this work, we adopted the version  2.2.1 of {\tt POSYDON} and Data Release 2.0.0
of its binary grids \citep{POSYDON_v2_zenodo}. These grids offer increased mass density in the mass interval $[4~ M_\odot - 14~ M_\odot ]$ compared to higher mass parameter space, providing enhanced mass---and therefore time---resolution (see Appendix \ref{sec:Appendix A}) to improve the accuracy of our synthetic spectra. The {\tt POSYDON} grids used to produce the stellar populations assume a solar metallicity of $Z = 0.014$ and solar abundances from \cite{2009ARA&A..47..481A}.

Our stellar populations include a combination of both single and binary stars. We assign the initial binary parameters by drawing 
from predefined distributions of primary mass ($m_1$), mass ratio ($q$), and orbital period ($P$). The primary star has an associated probability of being in a binary from the adopted binary fraction distribution, $P_{\mathrm{bin}}$. To determine whether the primary is assigned a companion or not we draw a random number $u$ from a uniform distribution over the interval $[0,1]$; if $P_{\mathrm{bin}} > u$ then it is evolved as a binary, while otherwise it is evolved as a single star.

To create our fiducial populations, our primary masses are drawn from a Salpeter IMF \citep{1955ApJ...121..161S} with a range of $0.1~ M_\odot$ to $150~ M_\odot$. The binary orbital period is drawn from a range of $\log P$ from -0.13 to 3.7, following the distribution from Equation 23 of \cite{2017ApJS..230...15M} which depends on the primary mass. 

For companion star masses, we adopt the probability distribution of the mass ratio from \cite{2017ApJS..230...15M} which incorporates a broken power-law distribution with different indices for small $q \in [ 0.1 - 0.3 ]$, $p_q \propto q^{\gamma_{\rm small}}$, and large mass ratios $q \in [ 0.3 - 1 ]$ $ p_q \propto q^{\gamma_{\rm large}}$, as well as the excess fraction $F_{\rm twin}$ of systems with a mass ratio $q > 0.95$. All three parameters of the mass ratio distribution ($\gamma_{\rm small}$, $\gamma_{\rm large}$, $F_{\rm twin}$) follow from 
\cite{2017ApJS..230...15M} and are both $m_1$ and $P$ dependent. Finally, the binary fraction of our fiducial population models also follows from \cite{2017ApJS..230...15M} which is mass dependent.

Once the initial parameters are assigned, each binary system or single star is evolved from the zero-age main sequence (ZAMS) through its subsequent stages of stellar evolution using the {\tt POSYDON} framework until a desired age. 
We produce 43 co-eval populations corresponding to starburst ages spanning from 1 Myr to 15.8 Gyr, in order to contain the current age of the universe, with an increment of 0.1 dex in log scale. Rather than setting the final ages for all stars in a population to be identical, we randomly draw end times from a flat distribution centered on that age, with a bin width optimized based on the effects of stochasticity and resolution limits of the binary grids; further details are provided in Appendix \ref{sec:Appendix A}. 

One caveat of our models is that {\tt POSYDON} data release 2 focuses on high-mass stars in binaries, with primary masses $>$4 $M_\odot$ (although companion star masses extend down to 0.5~$M_{\odot}$). We approximate the evolution of systems with less massive primary stars by including the contribution from binaries, but in this case simulating these low-mass binaries as two single stars on very wide orbits such that they never interact. For computational efficiency, the high ($m_1 >4 M_\odot$) and low ($m_1 < 4 M_\odot$) mass populations are modeled separately. By simulating these two populations separately, we also minimize the effects of stochasticity introduced by the Monte Carlo sampling method. For each of the high- and low-mass populations, we randomly draw half a million binaries and compute their integrated spectra. We then produce the final starburst spectral models by combining the spectra of the low- and high-mass populations, weighted according to the IMF, such that the resulting spectrum corresponds to a population with an initial stellar mass of $10^6 \, M_{\odot}$. 
Note that because our method takes a Monte Carlo approach, it allows for the introduction of stochasticity into our model spectra. \citet{2004A&A...413..145C} have shown that infrared emission in particular is sensitive to the model population size \citep[see also][]{2006A&A...451..475C, 2013NewAR..57..123C}. Because we simulate half a million systems for both our high-mass and low-mass stellar samples, we expect that these effects are small. Indeed, \cite{2023MNRAS.522.4430S} find that for clusters with stellar masses greater than $10^5 \, M_\odot$, the effects of stochasticity from binary populations are minimal; but a more complete study, which we leave for future work, is required to determine the impact of stochasticity on our binary spectral models (see Appendix~\ref{sec:Appendix C} for a comparison between our Monte Carlo approach and an alternative grid-based approach that adopts IMF weighting).


To extract stellar properties at specific times we use a combination of interpolation methods for our binary and single star systems. For single stars, {\tt POSYDON} interpolates grids of {\tt MESA} models using the equivalent evolutionary phases (EEPs) formalism of \cite{2016ApJS..222....8D}; stellar properties are derived as a function of time between the different models in the grids. For binary stars, analogous interpolation methods within {\tt POSYDON} have not yet been integrated \citep[but see][for work in progress]{2025ApJ...984..154S,2026arXiv260413604D}. Therefore, for any arbitrary binary we identify the nearest neighbor model run within the {\tt MESA} binary grid and interpolate the track in time to obtain the star's properties at the exact requested time. While this mapping procedure introduces some error into our results, the high model density of the binary grids within {\tt POSYDON}, especially for high mass stars whose properties vary rapidly with time, suggests the error introduced is minimal, typically bellow 1\% \citep{2023ApJS..264...45F}.

While we refer the reader to \citet{2023ApJS..264...45F} and \citet{2025ApJS..281....3A} for a complete description of how {\tt POSYDON} treats binary evolution, we highlight here some of the key components that are particularly relevant for the production of spectra. First,  by using {\tt MESA} {\tt POSYDON} is able to model the response of both donor and accretor (when the accretor is a non-degenerate star) throughout the binary's evolution including mass transfer phases.
For some types of binaries, a long phase of stable mass transfer can strip the star of its H envelope, leaving behind a hot, stripped He star. In {\tt POSYDON} the amount of mass accreted by the companion is regulated by its spin-up during mass transfer. As the accretor gains mass and angular momentum, it spins up. If the star reaches critical rotation, the mass stops being accreted and it leaves the system carrying away the specific angular momentum of the accretor. As a result, a large fraction of stable mass transfer systems evolve through highly non-conservative mass transfer, with a rapidly rotating accretor. 

In certain cases, especially when the donor star is a giant characterized by a flat entropy profile, the mass transfer phase can be unstable, leading to the H envelope of a giant star engulfing the binary. The so-called common envelope (CE) then is rapidly ejected through the shrinkage of the orbit, exposing a bare He star beneath in a tight binary with its companion. To model the post-CE outcome, we use the $\alpha_{CE} - \lambda_{CE}$ prescription (\cite{1984ApJ...277..355W}, \cite{1988ApJ...329..764L}), where $\lambda_{CE}$ is used to parameterize the binding energy of the donor's envelope and the  $\alpha_{CE}$ relates the fraction of the orbital energy used to unbind the envelope. In {\tt POSYDON} the parameter $\lambda_{CE}$ is being calculated from the structure of the donor during the onset of the CE phase, by defining a core-boundary defined at a specific H mass fraction. In our populations we adopt the values from \cite{2025arXiv250800186C}, that found that core-envelope boundary at $X = 0.3$ and $\alpha_{CE} = 1$ can better reproduce the observed distribution of galactic double neutron stars systems. If the binary has enough energy to eject the CE, we assume that the rest of the donor's envelope is removed through a second episode of stable mass transfer and it results in a stripped He-rich star with $X< 0.01$. This phase, too, has been suggested in previous work \citep{2019A&A...629A.134G} to be one of the key formation scenarios of binary stripped stars.

Systems that lack sufficient orbital energy to eject the envelope result in a stellar merger. Products of stellar mergers continue their evolution as single stars in {\tt POSYDON} \citep[for more details see section 5.2.3 of][]{2025ApJS..281....3A}. Stellar mergers produce massive stars that otherwise would not be present as the merger can exceed the lifetime of single stars with the same initial mass, appearing as blue stragglers \citep{1995ARA&A..33..133B,1990AJ....100..469M} in our populations. While the process of stellar mergers is still not fully understood, simulations suggest that most of the mass is likely retained, at least in the case of main sequence mergers \citep{2019Natur.574..211S}. Within {\tt POSYDON}, for any binary that merges, we calculate its stellar properties  such as total mass, core mass, and central abundances \citep[as described in][]{2025ApJS..281....3A}, and match it to the nearest similar single star, although any rotation gained or different core-envelope structure is not considered in the future evolution of the merger product.

When the primary reaches C depletion, we determine its final fate by using the core-collapse prescriptions of \cite{2020MNRAS.499.2803P}, and adopt a Maxwellian distribution for the supernovae (SN) kicks with a dispersion velocity of 265 km s$^{-1}$. Depending on the post-SN outcome, the binary can get disrupted, or continue its evolution as a detached system with a star and a compact object (CO) companion. {\tt POSYDON} models the subsequent evolution of the detached systems, including further episodes of mass transfer after the secondary overfills its Roche lobe. These interacting CO companion and star systems provide a new formation mechanism for binary stripped stars.

\begin{figure*}
                \centering
                \includegraphics[width=1\linewidth]{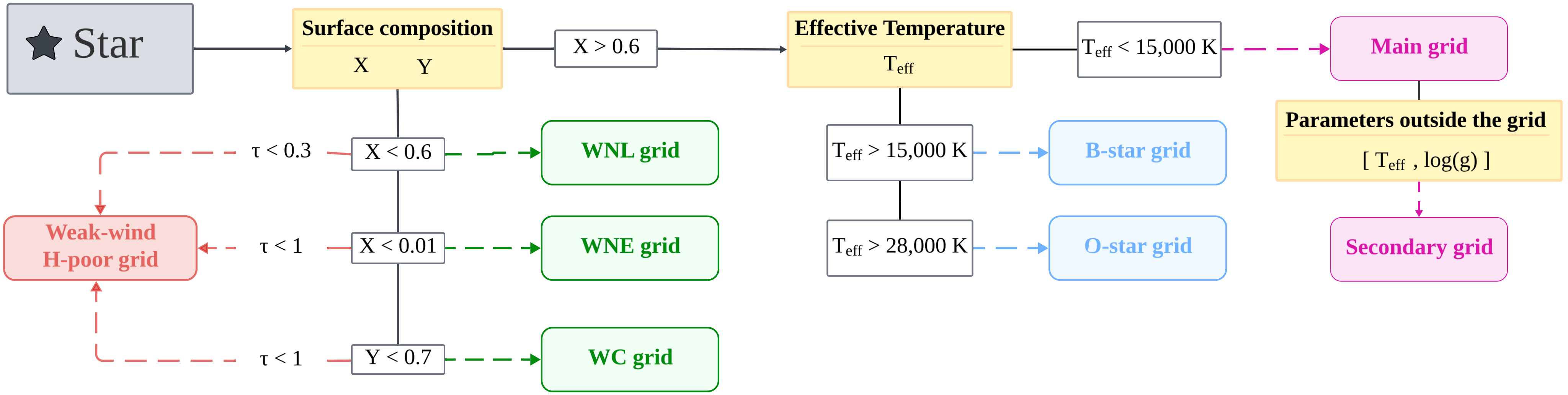}
                \caption{The flow chart outlining our procedure for identifying from which library each star's synthetic spectrum will be generated, based on its input parameters such as $log~g$, $T_{\rm eff}$ and $X_H$. Spectra for H-poor stars, with surface $X_H < 0.6$ like WR stars and stripped stars, are drawn from bespoke spectral libraries. For hot stars, we use either the {\tt OSTAR} or {\tt BSTAR} grids depending on $T_{\rm eff}$ \citep{2003ApJS..146..417L, 2007ApJS..169...83L}, while cooler stars use either the C3K library \citep{2014ApJ...780...33C}, or BOSZ library \citep{2024A&A...688A.197M}.}
                \label{flow_chart}
\end{figure*}

\subsection{The spectral synthesis } 



After we evolve our stellar populations we calculate the integrated flux at each wavelength as a sum of all the individual stars' spectra. For each star we infer its spectrum from a stellar library based on its stellar and surface properties. Since our populations include stars that span a wide range of temperatures and evolutionary phases we have stitched together a collection of spectral libraries, each of which covers a different parameter space of stellar properties, including effective temperature ($T_{\mathrm{eff}}$), surface gravity ($\log g$), and metallicity. Some libraries also include additional dimensions like the surface abundance of hydrogen or helium. Each of the libraries employed has an extended wavelength range spanning the ionizing regime to the far-IR and sufficiently broad stellar parameter coverage. Although our current models are only applicable for solar metallicity stellar atmospheres, we prioritized libraries in the literature that include multi-metallicity atmospheres, as they would be needed in future projects.

To assign a stellar library to each star in our population based on its stellar properties, we follow a library selection scheme as illustrated in Figure \ref{flow_chart}. The first step of the selection process is determining which stars are H-rich or H-poor. 
Stars with a surface hydrogen mass fraction of $X < 0.6$ are considered H-poor in our framework and are then matched to dedicated spectral libraries for either Wolf-Rayet (WR) stars \citep[based on the PoWR code;][]{2004A&A...427..697H, 2012A&A...540A.144S, 2015A&A...579A..75T} or to weak-wind H-poor spectra, (constructed with models from \cite{2023ApJ...959..125G}, developed using the non-LTE radiative transfer code CMFGEN \citep{1990A&A...231..116H, 1998ApJ...496..407H}). These two spectral grids differ since the WR spectra are characterized by strong emission and dense stellar winds, while the weak-wind H-poor spectra are constructed assuming lower wind mass-loss rates, leading to absorption line spectra. The distinction between the two libraries depends on the wind optical depth $\tau$, following \citet{2022A&A...661A..60A}. We describe in detail in Appendix \ref{sec:H-poor treatment} how we utilize $\tau$ among other stellar parameters to make the library selection for H-poor stars.

For stars with a surface hydrogen abundance $X > 0.6$, the second step in the selection process is based on stellar temperature which ultimately determines the star's spectral type. We have separate dedicated libraries for O and B stars that incorporate NLTE corrections using the TLUSTY models: {\tt OSTAR} for stars with $28{,}000~\mathrm{K} \le T_{\rm eff} \le 55{,}000~\mathrm{K}$ and {\tt BSTAR} for stars with $15{,}000~\mathrm{K} \le T_{\rm eff} < 28{,}000~\mathrm{K}$, adopted from the grids of \cite{2003ApJS..146..417L} and \cite{2007ApJS..169...83L} respectively. For all the remaining spectral types ranging from A--M stars ($T_{\rm eff} < 15,000~{\rm K}$), we use the C3K library of \cite{2016ApJ...823..102C} based on the ATLAS12 models \citep{1970SAOSR.309.....K,1993sssp.book.....K}, consisting of plane parallel LTE atmospheres over a range of several metallicities. Occasionally, the matching process fails for these libraries, either due to numerical reasons or, more rarely, the absence of a spectral model for that particular combination of stellar parameters, in which case we have additionally included a secondary spectral library, the BOSZ spectral library by \cite{2024A&A...688A.197M}, assigned to the systems that fail the primary matching. This grid is comprised from the APOGEE-ATLAS atmospheres for stars above $8000$ K and the MARCS atmospheres for stars cooler than $8000$ K. A complete list of the spectral grids and their characteristics are provided in Table~\ref{tab:spectral_grids}.

    \begin{table*}
    \centering
    \movetableright=-1.3in
    \resizebox{1\textwidth}{!}{%
    \begin{tabular}{|c|c|c|c|c|c|c|c|c|}
    \hline
    Grid Label &Library & \centering Wavelength (\AA) & Resolution & Surface H & $T_{\rm eff}~({\rm K})$ & $\log g$ $(cgs)$ & Metallicity \\\hline
    
    Main grid &C3K &	100--9,000  & R $\sim$ 3000  &	0.7381	&2500--46,000 	&-1.0--5.5	&[Fe/H]:-2.5--0.5  \\
    Secondary grid &BOSZ &	90--319,972  &	R $\sim$ 5000 &0.7381	&2800--16,000 	& 2.0--5.5	&[Fe/H]:-2.5--0.5 \\
    B-star grid &BSTAR2006 &	880--50,000 	& R $\sim$ 10,000 & 0.735 &15,000--30,000 	& 1.75--4.75	& $Z/Z_\odot$ : 0--2\\
    O-star grid &OSTAR2002	& 880--50,000  	& R $\sim$ 10,000& 0.735 & 27,500--55,000  & 3.0--4.75	& $Z/Z_\odot$ : 0.02--2 \\
    Weak-wind  H-poor grid &Gotberg23 & 110--20,000 
    & &0.01--0.7  & 30,000--150,000  & 4.0--6.0 & $Z = 0.006 $ \\\hline
    Grid Label &Library & \centering Wavelength (\AA) & Resolution & \centering Surface abundance & $T_{\rm eff}$ (K) & $R_t$ $(R_\odot)$ & Metallicity  \\\hline
    WNL grid &PoWR-MW-WNL & $ 5$--$8.4 \times 10^{7} $ & $R \sim 5000$ &H: 0.2, 0.5 &30,000--100,000  &1.0 - 50.1 &  $Z_\odot$  \\
    WNE grid &PoWR-MW-WNE &$ 5$--$8.4 \times 10^{7} $ & $R \sim 5000$ &He: 0.98 & 31,623--177,828 &1.0 - 50.1 &  $Z_\odot$ \\ 
    WC grid &PoWR-MW-WC &$ 5$--$8.5 \times 10^{7} $ &$R \sim 5000$ & He : 0.55, C : 0.4 &39,810--199,526 &0.31 - 39.8 &  $Z_\odot$  \\\hline
     
    \end{tabular}
    }
    \caption{\label{tab:spectral_grids} Summary of the spectral libraries and their parameters utilized in our spectral synthesis framework, together with the corresponding grid labels used in Figure \ref{flow_chart}.}
    
    \end{table*}

We highlight three additional caveats. First, stars that are transitioning from a post-AGB phase into a white dwarf (WD) exhibit a brief phase of evolution in which they are extraordinarily hot ($T_{\rm eff} \sim 10^5$ K) and bright ($L \sim 10^3~ L_\odot $). These  stars are often found surrounded by their  envelope material lost in previous phases of evolution, through which a fraction of their emitted ionizing radiation is expected to be absorbed. As we currently lack a prescription to account for this brief evolutionary phase, we defer a careful treatment of these systems for future work \citep[but see ][for an implementation]{2019AJ....158....2B}. We therefore artificially remove from our sample any stars that have $\log g > 6 $ and $\log L/L_\odot < 4 $, as well as stars that have depleted He in their core, defined by a central He abundance $< 0.01$, and also satisfy $\log g > 6 $. Second, we do not currently evolve the WDs found in our populations along their cooling tracks and their flux contribution is absent from our models. Finally, we do not incorporate any luminosity due to accretion into our models, such as the UV emission produced by accreting WDs or the X-rays produced by disks around neutron stars and black holes.

Once the appropriate library has been selected, we compute each star's spectrum by interpolating within the models of the selected spectral library. To do this we use library grids, where each grid point corresponds to a spectral model of an atmosphere with specific stellar parameters like $\log g$ or $T_{\rm eff}$, (see Table \ref{tab:spectral_grids} for others). For the interpolation, we utilize the Multi-dimension Spectral Grid \citep[{\tt MSG};][version 1.3]{2023JOSS....8.4602T}, a tool specifically designed for large-scale efficiency. {\tt MSG} employs multivariate interpolation, providing flexibility in handling spectral libraries of arbitrary dimensionality. Additionally, we use {\tt MSG}'s suite of tools to construct and refine the spectral library grids that are integrated into our code. {\tt MSG} requires spectral libraries that are regularly spaced in their input parameters to perform tensor product interpolation. However, gaps in the parameter space are common in the available spectral library data. {\tt MSG}'s strict dimensionality requirement can throw out some existing models or lead to occasional interpolation failures in regions where the grid has gaps. To avoid having these artificial ``voids", we fill up missing grid points with synthetic spectra generated through linear interpolation or, in some edge cases, extrapolation in the $\log g$ parameter.


One other rare failure mode can occur when one of the star's parameters falls outside the library grid, or if the spectral interpolation produces a flux with any negative values due to numerical artifacts produced by the interpolation (which can occur since {\tt MSG} calculates spectra using a cubic spline). As a last resort, in these cases the star is matched to the nearest spectral grid point.

The spectral libraries used in this work have different resolutions and wavelength ranges (see Table \ref{tab:spectral_grids}). In addition to spectral interpolation, MSG can resample the spectrum from a spectral library to a user defined wavelength grid. In this work  we used a wavelength range spanning from $110 \,\AA$ to $10 \, \mu m$ with a $ \Delta \lambda =  0.3 \, \AA$. For the libraries whose intrinsic wavelength coverage does not extend over this range, we extrapolated each spectrum using a blackbody approximation and reconstructed the spectral library to include our selected wavelengths. Finally, once we compute the individual spectra of all the stars in our population, we sum them all together to create the integrated spectrum of the entire stellar population.

\label{sec:spectral_libraries}

\label{sec:Methodology}

\section{POSYDON Binary Spectral Models}
\label{sec:binary_spectral_models}

Using our spectral synthesis framework we create the spectra of our single and binary starburst models\footnote{We provide the data used to generate the results described in this work available for download from Zenodo under an open-source Creative Commons Attribution license:\dataset[10.5281/zenodo.22923179]{\doi{10.5281/zenodo.22923179}}. This record includes the binary and single star starburst spectra for all ages and the \texttt{POSYDON} initialization files that were used to produce the populations.}.
In Section~\ref{sec:fiducial_pop}, we begin by describing our fiducial model and its time evolution. In Section~\ref{sec:contribution}, we break down the contributions of different stellar populations to the overall spectra. Then, in Section~\ref{sec:H-poor_stars} we present the evolution and types of WR stars in our fiducial models as well as the formation and contribution from stripped stars. Finally, in Section~\ref{sec:ionizing_radiation} we discuss the ionizing output from our fiducial model. 

\begin{figure*}
    \centering
    \includegraphics[width=1\linewidth]{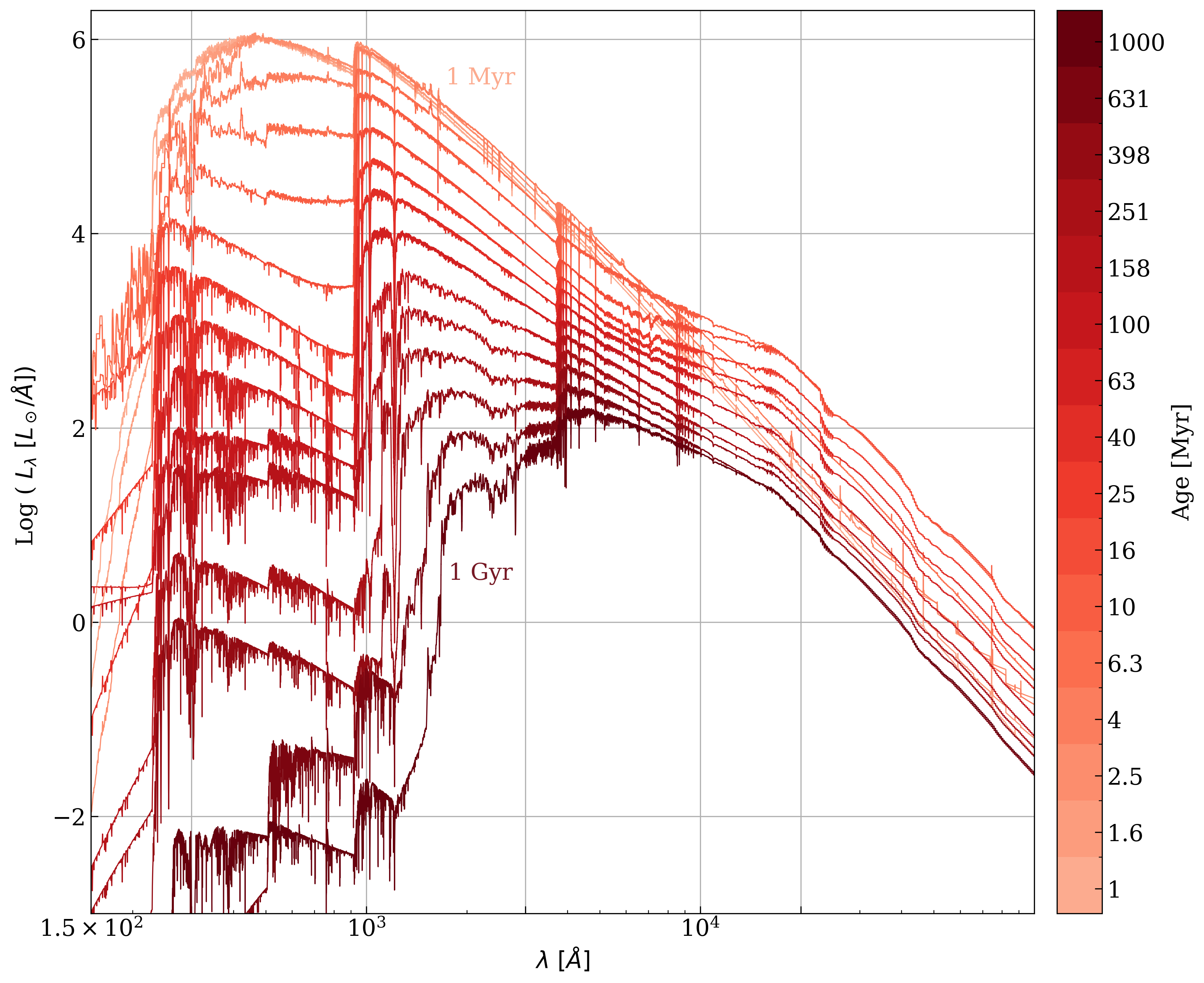}
    \caption{ The spectral energy distribution of our fiducial stellar population model over time including binaries. The populations span a range of ages from 1 Myr to 1 Gyr. 
    }
    \label{fig:Time_evolution}
\end{figure*}

\subsection{Fiducial Population Spectra}
\label{sec:fiducial_pop}

In Figure \ref{fig:Time_evolution}, we illustrate the evolution of our synthetic spectra for a coeval fiducial population as a function of age. The spectral evolution of stellar populations has been comprehensively described elsewhere in the literature \citep{1999ApJS..123....3L,2003MNRAS.344.1000B,2005MNRAS.357..945G}. Nevertheless, we provide a brief description here of the main components exhibited in Figure~\ref{fig:Time_evolution} for completeness, with an emphasis on the impact of binarity. In the first few Myr after a stellar population forms, massive stars dominate the emission: O-type stars and WR stars produce the bulk of the Lyman continuum (LyC) radiation ($\lambda < 912$ \AA), while O- and B-type stars dominate the UV continuum ($\lambda > 912$ \AA). As the population ages {($\simeq 10$ Myr), massive stars evolve off the MS to become red supergiants. As a consequence, we simultaneously see a decline in the UV accompanied by an increase in the infrared output ($\lambda \gtrsim 10^4$ \AA).  

When massive stars cross the Hertzsprung Gap, their envelopes expand and, for the subset that are formed in binaries, start filling their Roche-Lobes (RL), triggering episodes of mass transfer. These binary interactions create a population of stripped stars. We can see this transition most prominently in the ionizing regime, which evolves from showing strong emission features from hot WR stars to spectra with deepening absorption features characteristic of the weaker winds in the atmospheres of the exposed He-cores of binary stripped stars. 

Over time (4 Myr -- 158 Myr in our simulations) the Lyman break becomes more prominent as the OB-type stars evolve past the MS and the ionizing continuum ($ \lambda < 912$ \AA) declines. The existence of WRs and stripped stars boosts the flux in the extreme-UV, but the overall continuum below 912 Å is still strongly reduced compared to the youngest ages. We also see the hydrogen absorption lines deepening over that period, reflecting the aging stellar content that is starting to be dominated by B-type and A-type stars and later by late A and early F-type stars.

\begin{figure*}
        \centering
        \includegraphics[width=1\textwidth]{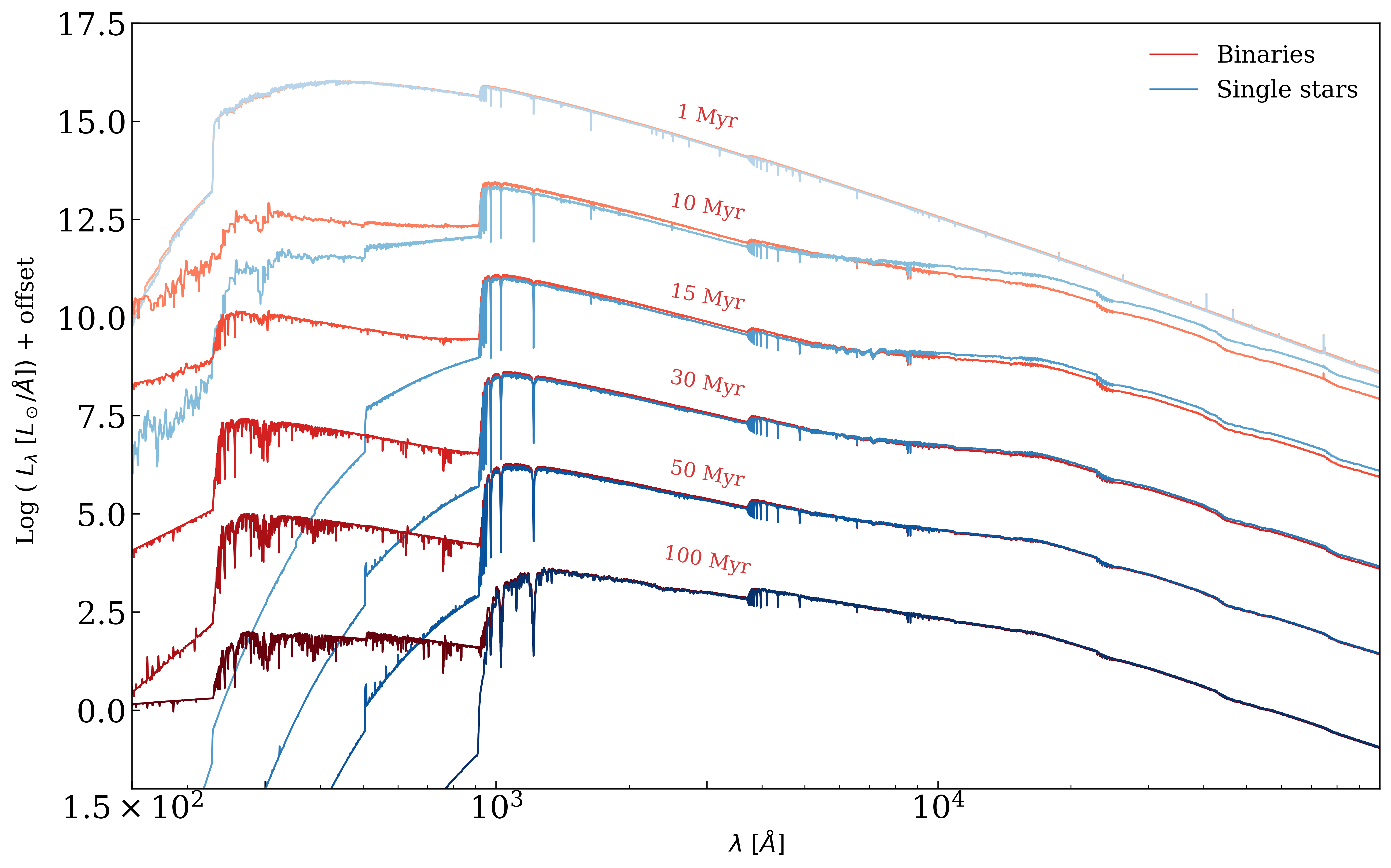}
        \caption{ We show the spectra from binary and single star populations over several ages ranging from 1 Myr to 100 Myr. The binary spectra are plotted in the shades of red, while single star populations are in blue colors. Shortly after 1 Myr, we see a steep drop in the UV radiation from the single stars populations, in stark contrast to binary populations, which continue to emit hard ionizing radiation over time due to the presence of stripped stars. 
        }
        \label{fig:Single_star_comp}
\end{figure*}

To isolate the effects of binary evolution in the spectral output of our populations, we evolve comparison populations containing only single stars following the same IMF and create spectral models that correspond to a population with a total mass of $10^6 M_\odot$. In Figure \ref{fig:Single_star_comp} we show the results from our single and binary star populations at specific ages: 1, 10, 15, 30, 50, and 100 Myr. This comparison underscores the differences in evolutionary behavior that is inherent to binary populations.

While there are some small differences between the models in the visible and the infrared wavelengths, the most significant divergence of the models occurs in the FUV, at ages $\gtrsim$15 Myr. Our results are in agreement with previous studies demonstrating that binary populations produce bluer spectra \citep{2017PASA...34...58E, 2019A&A...629A.134G}. Young single star populations are able to produce equivalent amounts of ionizing flux within the first few Myr, but as early as 10 Myr we see a steep decline that is not present in our binary populations. The sustained LyC luminosity is driven by intermediate ($4~ M_\odot < M_{\mathbf{ZAMS}} < 20~ M_\odot$) stripped enveloped stars, which produce the copious amounts of extreme UV radiation that are absent from the single star models. 

In addition, we observe that at 10 Myr, 15 Myr and 30 Myr the binary populations produce less flux at wavelengths larger than $5000$ \AA. We attribute this difference to the reduced number of giant stars in the binary models, as envelope stripping prevents a number of stars from evolving to cool supergiant phases, and at older ages from ascending the AGB. Since their high luminosity and low effective temperatures make giant stars the dominant contributors at longer wavelengths, envelope stripping can reduce their contribution \citep[by as much as 30 \%;][]{2017PASA...34...58E}, leading to the flux differences observed in our models.

\begin{figure*}
    \centering
\textbf{}    \includegraphics[width=1\linewidth]{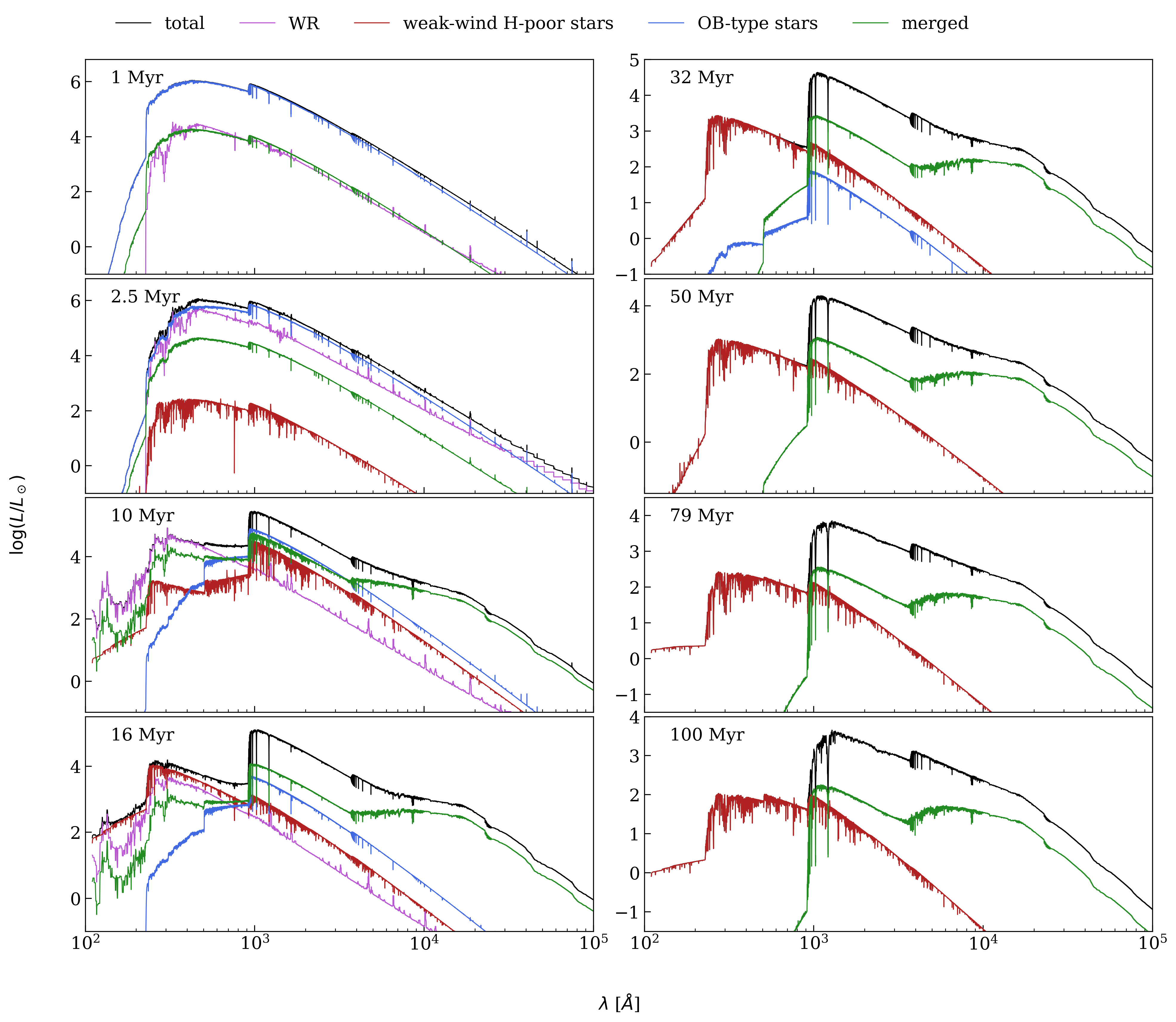}
    \caption{Breakdown of spectral contributions of ionizing and UV-bright sources
    for several ages ranging from 1 Myr to 100 Myr. We display with different colored lines the contributions from binaries populations such as WR (purple), O\textbf{B}-type stars (blue), stripped stars (red) and merged stars (green). We also include the flux of the overall population (black). For populations younger than 15 Myr the WR populations dominate the production of UV photons and after 15 Myr the He-rich stripped stars from binary interaction dominate the ionizing flux and ultimately become the sole contributor. }
    \label{fig:multi-contribution}
\end{figure*}

\subsection{Contributions from Sub-Populations}
\label{sec:contribution}

While Figure \ref{fig:Time_evolution} presents the aggregated flux of the entire population, in Figure \ref{fig:multi-contribution} we separate the contributions of key binary sub-populations and spectral types over time. We follow the evolution of hot and UV-bright emitters: the WR stars (magenta), the weak-wind H-poor stars (red), and OB-type stars with $ T_{\rm eff} \geq 28,000~{\rm K}$ (blue) from the overall population. We also show separately the contribution of stellar mergers (green) which is an evolutionary classification rather than a spectral one. We note that this category is not mutually exclusive; a merged star may simultaneously satisfy the criteria for an OB-type or WR star and thus contribute to the flux of both. Finally, we have not included the contribution of stars that do not fall under any of these specific categories, the B-type ($ T_{\rm eff} \leq 28,000$) and later stars that may produce significant fractions of the overall radiation, especially at later times. Therefore, the colored lines in Figure~\ref{fig:Time_evolution} do not equal the black, ``total'' line which represents the entire stellar population spectrum.

At very early ages ($\lesssim 10~{\rm Myr}$) the majority of the UV emission is produced by WR stars and OB-stars, as seen in the top three left panels of Figure \ref{fig:multi-contribution}. Hot OB-type stars, dominate the UV and LyC until 2.5 Myr, after which the WR stars start to become the dominant source of ionizing photons. At 16 Myr we see the transition of the extreme UV production from being WR dominated to being stripped-star dominated. In subsequent ages ($> 16~{\rm Myr}$), the the weak-wind H-poor stars are the sole sources of hard ionizing radiation, a result that is also evident from the comparison with the single star population shown in Figure \ref{fig:Single_star_comp}. As the stars continue to evolve, the underlying binary stripped star population shifts to lower masses reducing their overall luminosity output of the weak-wind H-poor stars. After 100 Myr, the UV production has been reduced by two orders of magnitude, but high energy photons are still being produced by binary stripped stars.

Another important contribution comes from products of stellar mergers. We distinguish between two different types of mergers in our populations that have their unique imprint on the spectra of mergers. The first type are binaries that merge when both stars are on the MS, and the outcome of the merger is a less evolved MS star with an increased mass that equals the total mass of the binary at the time of the merger, which appears as a blue straggler. These events are able to produce very massive stars and are responsible for a production of substantial radiation in the LyC. In Figure \ref{fig:multi-contribution} we can see the products of these MS mergers in the spectral distribution of mergers, which indicates the presence of both O stars and WR stars. Notably, the O-type star contribution in the 32 Myr panel of the figure is also the result of these mergers.

The second type involves mergers that were triggered from unstable mass transfer of an evolved donor onto a MS companion. We model these mergers using a prescription \citep[more details in][]{2025ApJS..281....3A} that engulfs the dense core of the donor, by an envelope that now includes the H-rich material of the companion star. The resulting star is a cool, extended giant. The presence of giant stars can been seen in the long wavelength end of spectral distribution where merger spectra have the characteristic bump attributed to stars ascending the giant branch.

Overall, we see that the contribution from stellar mergers in the UV decreases over time. At 1 Myr Figure~\ref{fig:multi-contribution} shows that the overall luminosity produced by mergers is equivalent to the luminosity produced by the entirety of the WR subpopulation. As time goes on, their contribution diminishes, but may still account for a significant fraction of the overall luminosity. 

\subsection{H-poor stars }
\label{sec:H-poor_stars}

The ionizing radiation of young stellar populations (($  \gtrsim 3 \mathrm{Myr}$)) is dominated by H-poor stars. In our populations we distinguish between WR stars and weak-wind H-poor stars (see Appendix \ref{sec:H-poor treatment} for details about how we make this distinction).

At very early ages ($ \lesssim 10 \mathrm{Myr}$) WRs are the dominant producers of UV radiation. In Figure~\ref{WR-stars}, we show the number of WR stars that are present in our fiducial model. According to the evolutionary scenario offered by \cite{1983ApJ...274..302C}, WRs are the descendants of massive O-type stars which first evolve to become a member of the WN subtype upon losing their H envelopes through strong stellar winds, exhibiting strong emission lines of helium and nitrogen. If they can maintain their strong winds they will evolve into a WC or WO subtype, characterized by helium, carbon, and oxygen emission lines \citep{2014A&A...564A..30G}. These subtypes correspond to different stages of nucleosynthetic processing that are exposed as the stellar envelope is being lost by winds; WN show the products of the CNO cycle in their atmospheres while the WC shows evidence of triple-$\alpha$ reaction products of He-burning. We further differentiate between the early-type WN stars that are hydrogen free as WNE, and the late-type WN stars that contain hydrogen as WNL.

In our model, we assign WR spectral subtypes following the criteria outlined in Appendix \ref{sec:H-poor treatment}. This evolutionary scheme agrees with the results from our population, as Figure~\ref{WR-stars} shows that the peak of WN stars precedes the peak of WC. We also see that WC stars, which require more evolved progenitors, are the first to disappear from our population and are only present for the first 10 Myr. 

Binary interactions are expected to affect the formation and the overall population of WR stars, as mass transfer provides an additional mechanism for removing envelope material. In Figure~\ref{WR-stars}, we include a comparison with our single star population model as dashed lines, where the differences introduced from binary interactions become apparent. In the first few Myr the number of WRs from the two different populations is equivalent, as these WRs are massive stars ($ M_{\text{ZAMS}} \gtrsim 20 M_\odot$) that can become stripped from winds alone. 
Binary interactions begin to play a role in the total number of WR stars produced after 3.2 Myr where mass stripping more than doubles the number of WR stars present in binary populations compared to single-star populations; the majority of which are of WNL type. The effects of binaries are two-fold: 1) mass stripping causes stars to enter their WR phase earlier in their evolution, extending the lifetime of the WR phase and 2) stellar mergers of lower mass stars produce WR stars at later times, when none are produced in single star populations. Binary interactions (through both envelope stripping via mass transfer and stellar mergers) extend the presence of WR stars to 17.5 Myr, in contrast to the single star populations where WRs are not produced after the first 10 Myr. 
\begin{figure}
    \centering
    \includegraphics[width=1\linewidth]{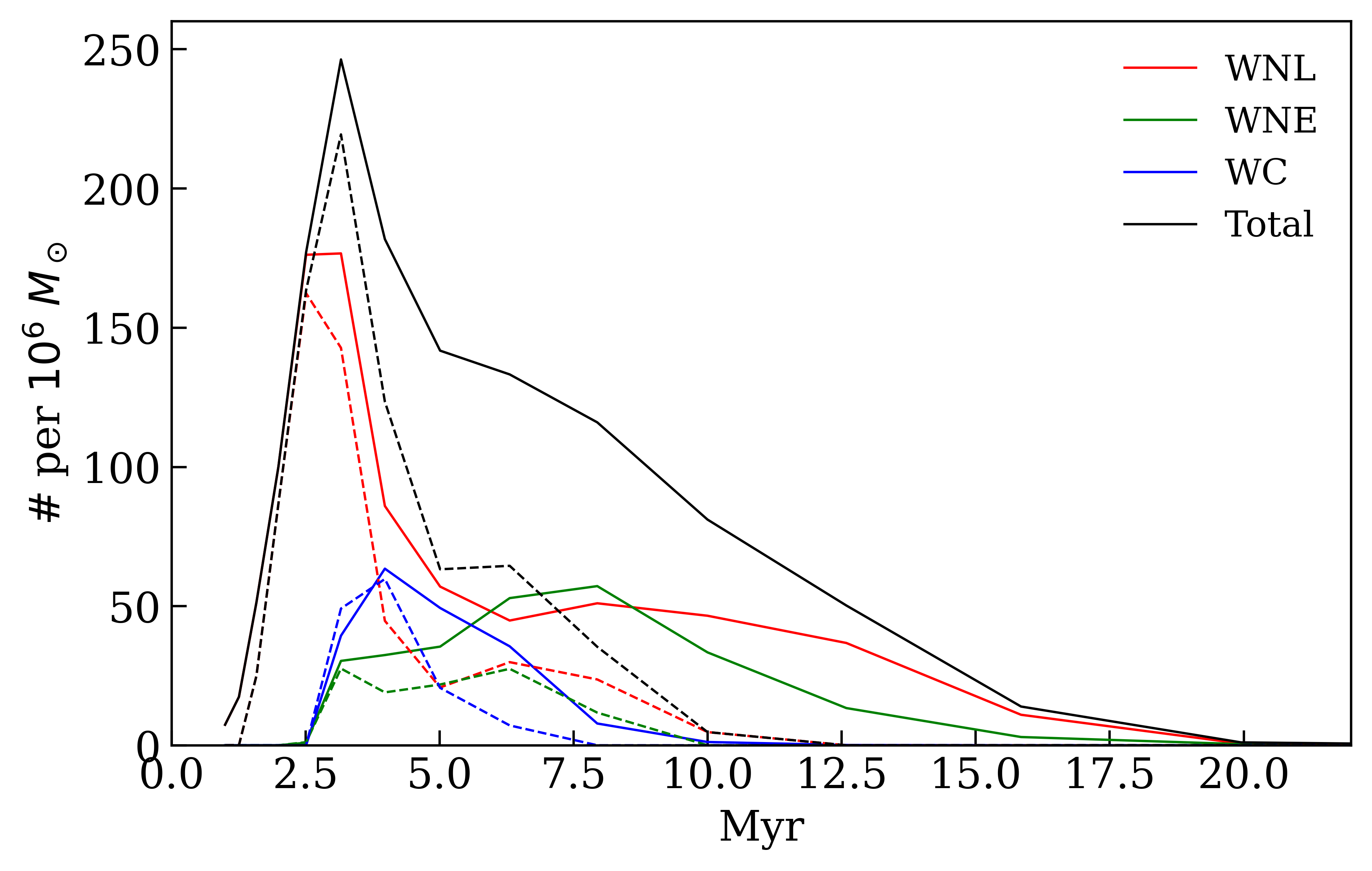} 
    \caption{The number of WR and their species in a starburst fiducial model as a function of time. The dashed lines represent the numbers from single star populations. Binary populations are able to produce and sustain relatively large numbers of WR stars over longer periods compared with the single stars.}
    \label{WR-stars}
\end{figure}

Binary stripped stars that are often assigned to the weak-wind H-poor star spectra, start to become important after 10 Myr, as stars with masses less than $18~ M_\odot$ begin to interact with their companions after evolving off the main sequence. Although binary stripped stars represent only a small fraction of the total stellar population, Figure~\ref{fig:multi-contribution} shows that they account for the majority of the emitted ionizing radiation in older binary populations \citep[in agreement with][]{2019A&A...629A.134G}.

However, binary stripped stars do not comprise a homogeneous class, as
stars can expel their H envelope through different mass transfer processes.
If the mass transfer is initiated during the MS phase, these stripped stars are labeled as Case A. Stars that fill their RL when they cross the Hertzsprung gap (HG), are labeled as Case B, while those that initiate mass transfer during and after core He burning are labeled as Case C. The cases above refer only to outcomes of stable mass transfer. If mass transfer becomes dynamically unstable then the binary enters a CE phase. We therefore define a fourth channel, the post-CE channel, in which the expulsion of the CE results in a system with a stripped-envelope star ($X < 0.01$) and MS star. Finally, we identify a fifth population of binary stripped stars that are produced when the secondary, the less massive star in the binary, is eventually stripped by stable or unstable mass transfer onto a compact object. For this formation channel we do not differentiate between the different mass transfer cases as we did for the primary stars. 

In Figure~\ref{stripped}, we show the time evolution of the instantaneous number of binary stripped stars present in our fiducial population model, along with the contributions from each formation channel. We find that around 58\% of binary stripped stars are produced from stable case B mass transfer. During H-shell burning, stars expand significantly and are able to transfer substantial mass to their companion on a thermal timescale before detaching. Therefore, stable Case B mass transfer occupies the largest area in the parameter space of the POSYDON grids for the systems with initial primary masses ($m_1$) between $4~ M_\odot - 18~ M_\odot$. The number of case-B binary stripped stars increases with time as their progenitors are more heavily weighted by the IMF.

Case A mass transfer, on the other hand, contributes a smaller fraction to the underlying binary stripped star population, accounting for approximately 15\%. In Case A systems, mass transfer begins while the donor is still undergoing core-hydrogen burning. Although few binaries undergo stripping through Case A mass transfer, since this
phase of mass transfer proceeds on the donor’s nuclear timescale, their contribution to the overall number of binary stripped stars at any one time is non-negligible. Indeed, we find that nearly half of the systems that are classified as Case A in any given population age
are actively undergoing mass transfer, while the remaining systems are detached. These systems typically retain moderate surface hydrogen mass fraction of $ 0.3 < X \le 0.6$ and have a lower $T_{\rm eff}$, than other stripped-stars. For this reason, the contribution of Case A binary stripped stars to the ionizing flux of the integrated spectra is disproportionately small relative to their number fraction in the stripped-star population.

The post-CE channel accounts for roughly 11\% of the overall population of binary stripped stars. However, we note that the contribution of this channel is the most uncertain, as CE remains a poorly constrained phase of binary evolution. We expect that changing key parameters in our CE prescription, especially the $\alpha_{\rm CE}$, will affect the production of binary stripped stars through this channel. 

Although we include the case C mass transfer as a distinct formation channel, we find that it produces a negligible number of systems and thus is not present in Figure \ref{stripped}. These stars tend to have shorter lifespans compared with the other binary stripped star production channels and are anyway more sensitive to numerical artifacts introduced by the finite resolution of our binary grids (we discuss the impact of grid resolution in more detail in Appendix~\ref{sec:Appendix A}). 

Finally, for systems that are able to survive a possible disruption event after the primary goes through core collapse, the remaining star is likely to eventually interact with its compact object companion. Binary stripped stars formed from secondaries are found to account for $\simeq$16\% of the total stripped star population, and are typically formed through Case B mass transfer. Their formation however, and the contribution of this channel is expected to be sensitive to the kicks from the adopted SN prescription.

\begin{figure}
    \centering
    \includegraphics[width=1\linewidth]{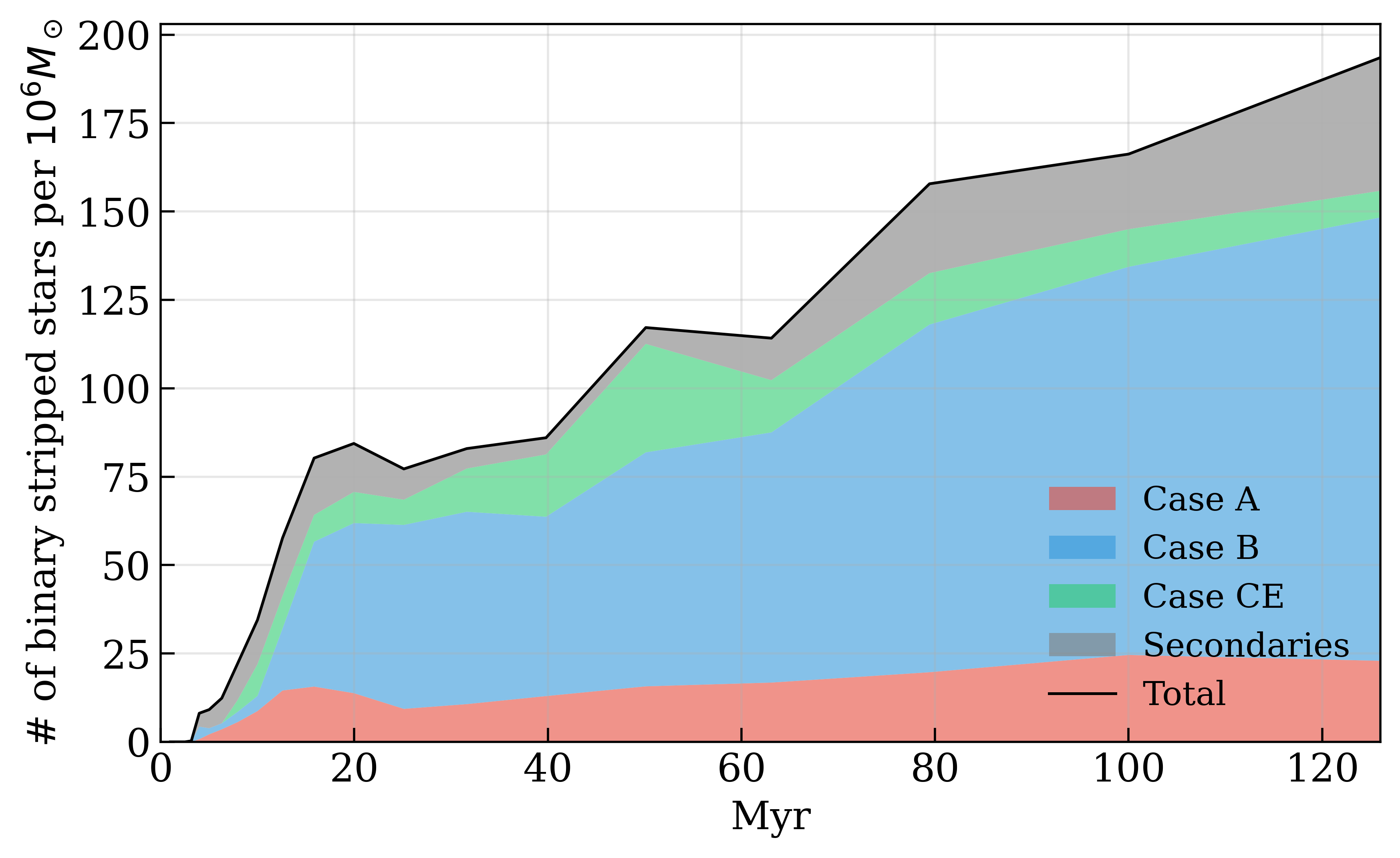}
    \caption{The number of binary stripped stars that are present in our fiducial population model as a function of time. We break down the formation channels and compare them with the total amount of binary stripped stars. These formation channels are binary stripped stars formed from stable case B mass transfer, stable case A mass transfer and stripped stars from post CE systems. We also include binary stripped stars formed from the secondary star. The majority of stripped stars come from the stable case B mass transfer channel. Stable case A and unstable mass transfer make up roughly 26 \% of the overall stripped star population, while binary stripped secondary stars account for approximately 16 \%.}
    \label{stripped}
\end{figure}

\subsection{Ionizing Radiation}
\label{sec:ionizing_radiation}


We focus on the emission rates of \text{H}, \text{He}, and $\text{He}^+$ ionizing photons, as the prediction of these species is important not only for modeling emission lines of young star-bursty ensembles but also in the context of cosmic re-ionization. We calculate the emission rate of ionizing photons $Q_i$, where $i$ corresponds to a specific species, by integrating from the minimum wavelength (typically 110 \AA, see Table \ref{tab:spectral_grids}) of the output flux to the ionizing edge of each species, $\lambda_{i, \rm edge}$:
\begin{equation}
    Q_i = \frac{1}{h c} \int_{\lambda_{\rm min} }^{\lambda_{\rm edge}} \lambda~ L(\lambda)~ d \lambda,
\end{equation}
where $c$ is the speed of light and $h$ is Planck's constant. The wavelength corresponding to the ionizing edges of the selected species are: 912 \AA, 504 \AA, and 228 \AA, for \text{H}, \text{He}, and $\text{He}^+$, respectively. 

In Figure~\ref{Ionizing radiation} we show the emission rate for both our fiducial population (solid lines) and the single-star-only population (dashed lines) as a function of time after a starburst of $10^6 M_\odot$. At young ages ($\leq$ 5 Myr), the single and binary populations produce similar ionizing fluxes. This result is consistent with the presence of massive MS O-type stars in both populations, as well as their similar numbers of WR stars at these ages as shown by Figure \ref{WR-stars}.

After 10 Myr the stripped-enveloped, He-rich stars in our populations are able to maintain substantial production of ionizing photons. Although the rate of ionizing photon production continuously decreases over time, it is characterized by a much shallower slope than what is seen in the single star population. 

The emission rate of the $\text{He}^+$ ionizing photons presents the biggest discrepancy between binaries and single stars. This difference is due to the vastly different spectral contributions after $\simeq20$ Myr, as, at these late times, only binary stripped weak-wind H-poor stars are able to produce in significant numbers the high energy photons that can fully ionize He. These results demonstrate that $Q_{\text{He}^+}$ is the most sensitive emission rate with regards to the effects of binary evolution and by extension the properties and formation channels of the underlying population of binary stripped stars. This result agrees with the study by \citet{2020MNRAS.495.4605S} of the sensitivity of the \ion{He}{2} $1640$ \AA\ emission line on the binary fraction, which directly impacts in the total number of produced binary stripped stars. 

\begin{figure}
    \centering
    \includegraphics[width=1\linewidth]{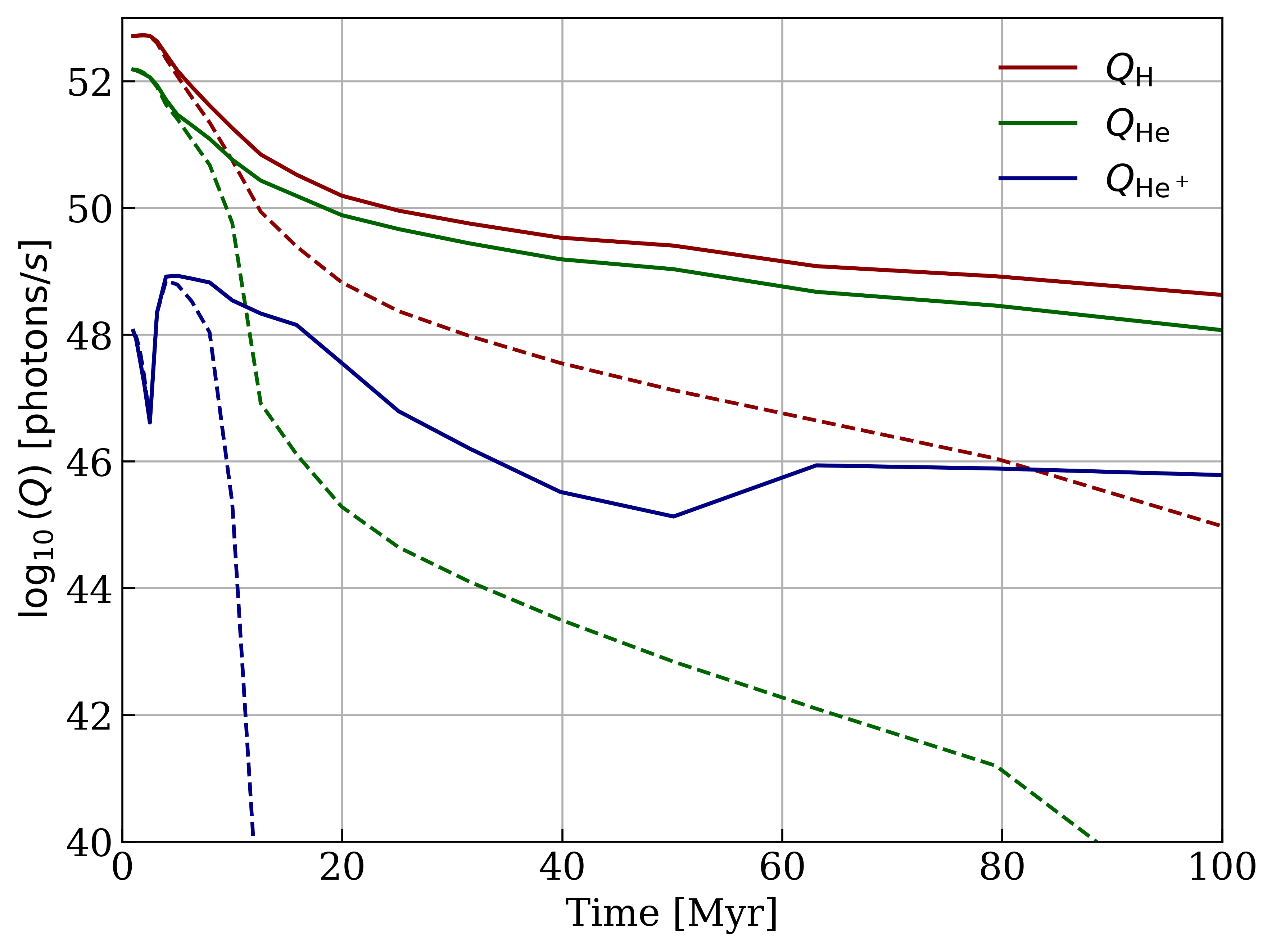}
    \caption{The emission rates of ionizing photons from binary and single star populations as a function of time. The solid lines correspond to the emission rates produced by the binary population and the dashed lines from only single star population. }
    \label{Ionizing radiation}
\end{figure}

\label{Spectral_models}   
\section{Comparison with Literature Spectral Models}
\label{sec:Model_comparison}

\begin{figure*}
    \centering
    \includegraphics[width=1\linewidth]{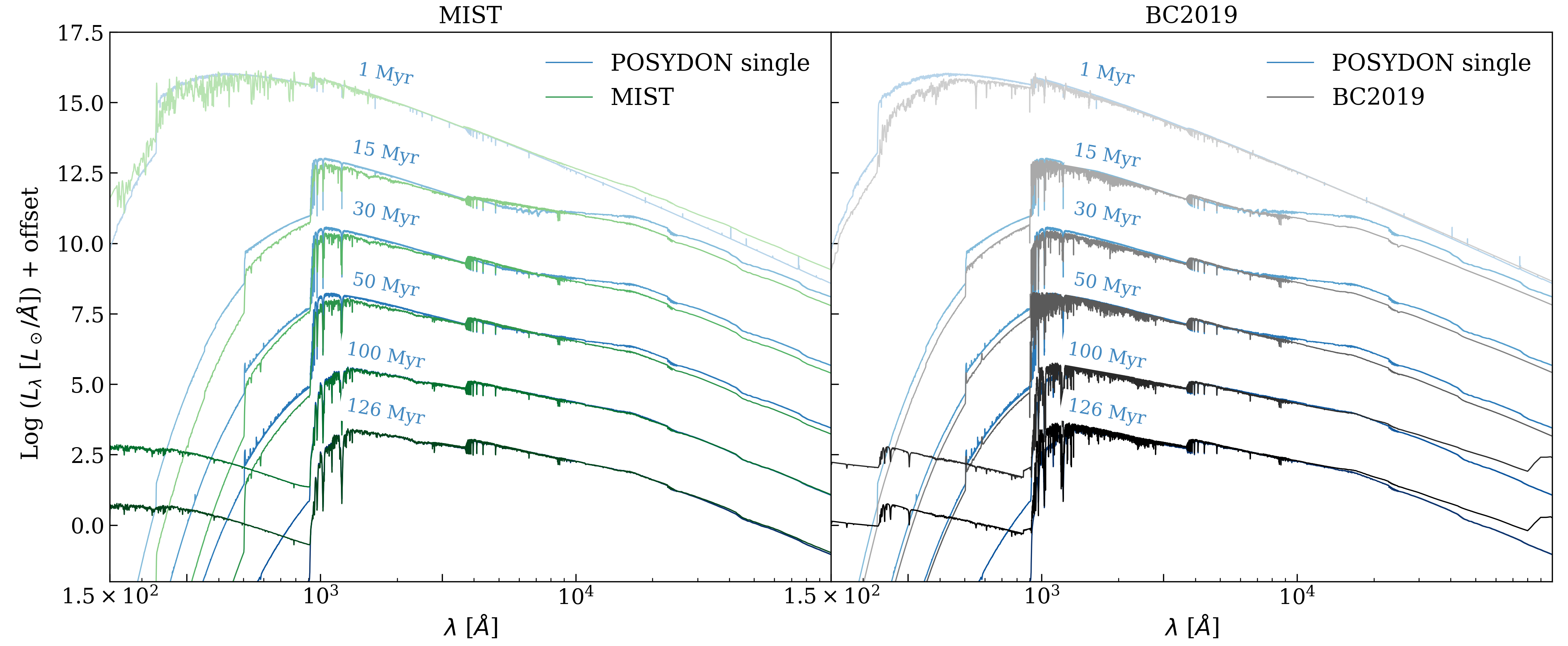}
    \caption{Spectral models of single-star populations at several ages. In the left panel, we compare
 the {\tt POSYDON} single-star models with spectral models produced using the MIST evolutionary tracks. In the right panel, we compare the {\tt POSYDON} single-star models with the BC2019 models.}
    \label{fig:single_MISTvsPARSC}
\end{figure*}

Our work builds off the extensive body of spectral synthesis models in literature.
We begin by focusing on single star populations, which serve as the benchmark to assess any differences in our stellar tracks and libraries compared with literature models, before we move on to examining the differences caused by binary evolution.

We start by comparing our single-star models with the MIST stellar evolutionary tracks evolved through {\tt MESA} from \cite{2016ApJS..222....8D} and \cite{2017ApJ...838..159C}. We compute the spectral output of the MIST models at several ages by using {\tt FSPS} \citep{2009ApJ...699..486C,2010ascl.soft10043C}. Both our models and {\tt FSPS} are based on {\tt MESA} stellar evolution tracks and use C3K as the primary spectral library.
However, for hot, massive stars {\tt FSPS} incorporates the WM-BASIC \citep{2012ascl.soft04001P} spectra (compared to our use of TLUSTY spectra), while for the WR stars, {\tt FSPS} uses CMFGEN WN and WC spectra from \citet{2002MNRAS.337.1309S}. We produce starburst models with {\tt FSPS} corresponding to the same ages as our own models and select a Salpeter IMF with a mass interval of [$0.1~ M_\odot$ - $150~ M_\odot$]. Our matching choice of initial conditions and spectral libraries for most stars (as a reminder, we use PoWR models for WR stars, not CMFGEN), allows for a direct comparison of the two stellar evolutionary models. 
We overplot our single star models and MIST models in the left panel of Figure \ref{fig:single_MISTvsPARSC}. 

We highlight three key differences between these models. First, we see an uptick in the LyC ($\lambda < 912$~\AA) for the MIST models at ages above 100 Myr and 126 Myr. We attribute this difference to the presence of WD and hot He-rich central stars of planetary nebula in the MIST models. Since we are removing these systems from our models, our single star populations are missing these sources of ionizing radiation. Second, our models produce somewhat more flux at all UV wavelengths from 15 Myr to 50 Myr. While we do not know the exact source of this emission, we speculate that it could be due to differences in the choice of underlying stellar parameters used to evolve stars within {\tt POSYDON}. We looked into potential sources of these differences by comparing the choices made for key parameters in the MESA simulations. One of these parameters in particular, convective coreovershooting, stood out.

Both MIST and POSYDON adopt a convective overshoot scheme of the exponential decay as parameterized by \citet{2000A&A...360..952H}.  $f_{ov}$ is a free parameter in this formalism, that determines the extent of the overshoot mixing. MIST adopts a constant value of $f_{ov} = 0.016$ for core overshooting, while in the POSYDON models the parameter is mass dependent and has a value of $f_{ov}= 0.0415$ for initial mass above $8~ M_\odot$ and for masses below $4~ M_\odot$ uses the same value as the MIST models. For stars with initial masses between $4~ M_\odot$ and $8~ M_\odot$, there is a smooth transition between these two values. This could potentially explain why at 100 Myr and 126 Myr the spectral outcomes of both models seem to overlap over these wavelengths, as stars of masses $M < 4~ M_\odot$ dominate the spectral distribution of populations at these ages.
Finally, our models exhibit an uptick in the \address{IR} flux at $1~ \mu m $ relative to MIST for ages younger than 100 Myr, while the models appear to agree at ages 100 and 126 Myr. Although these deviations may also be explained by the differences in the assumed stellar physics, we find that the population synthesis method itself can introduce variations in the IF. We explore this further in Appendix~\ref{sec:Appendix C}, where we compare the stochastic sampling and track interpolation used in our models with the IMF-weighting approach adopted by FSPS.

In the right panel of Figure~\ref{fig:single_MISTvsPARSC} we compare our models to the BC2019 updated version \citep{2019MNRAS.490..978P} of the GALAXEV code \citep[][]{2003MNRAS.344.1000B}. The BC2019 models use the PARSEC stellar tracks from \cite{2012MNRAS.427..127B,2015MNRAS.452.1068C}. For MS stars they use the stellar libraries from \citet{2015MNRAS.452.1068C} and include a treatment for their WR stars as described in \citet{2019MNRAS.490..978P}, where they utilize the PoWR WR models \citep{2002A&A...387..244G,2004A&A...427..697H,2012A&A...540A.144S, 2015A&A...579A..75T}.

The spectral differences between our models and the BC2019 models largely mirror those differences when compared to the MIST models; our models show increased emission at wavelengths shorter than $912$~\AA\ and again at infrared wavelengths longer than 1 $\mu$m for ages younger than 100 Myr. However, contrary to the MIST models, the UV emission between the Lyman and Balmer limits is largely consistent between our models and the BC2019 models, at least for ages 50 Myr or younger.

In Figure \ref{fig:MIST_PRSC_comp} we compare our fiducial binary population at 
1, 15, 30, 50, 100 Myr with the corresponding single star models from MIST and BC2019. The figure shows a similar behavior with that observed from the comparison to our own single star models: The binary models produce significantly enhanced amounts of UV radiation at wavelengths shorter than 912~\AA\ for every age older than our 1 Myr model. For the age of 100 Myr, the MIST and BC2019 produce ionizing radiation from WDs and post-AGB stars, we see that this contribution is comparable to, and in some cases exceeds, the ionizing output from stripped stars in our binary models for wavelengths less than $200$~\AA. 

The differences we see in our spectral models at wavelengths $1000 - 3000$~\AA\ and larger than 1 $\mu$m, also appear when comparing our single star models to the BC2019 and MIST models. Since this behavior is also present in the single star populations, we conclude that it cannot be attributed principally to the effects of binary interactions but rather to differences in the underlying assumptions adopted between the {\tt POSYDON} stellar physics and that of other models in the literature.

\begin{figure*}
    \centering
        \includegraphics[width=1\textwidth]{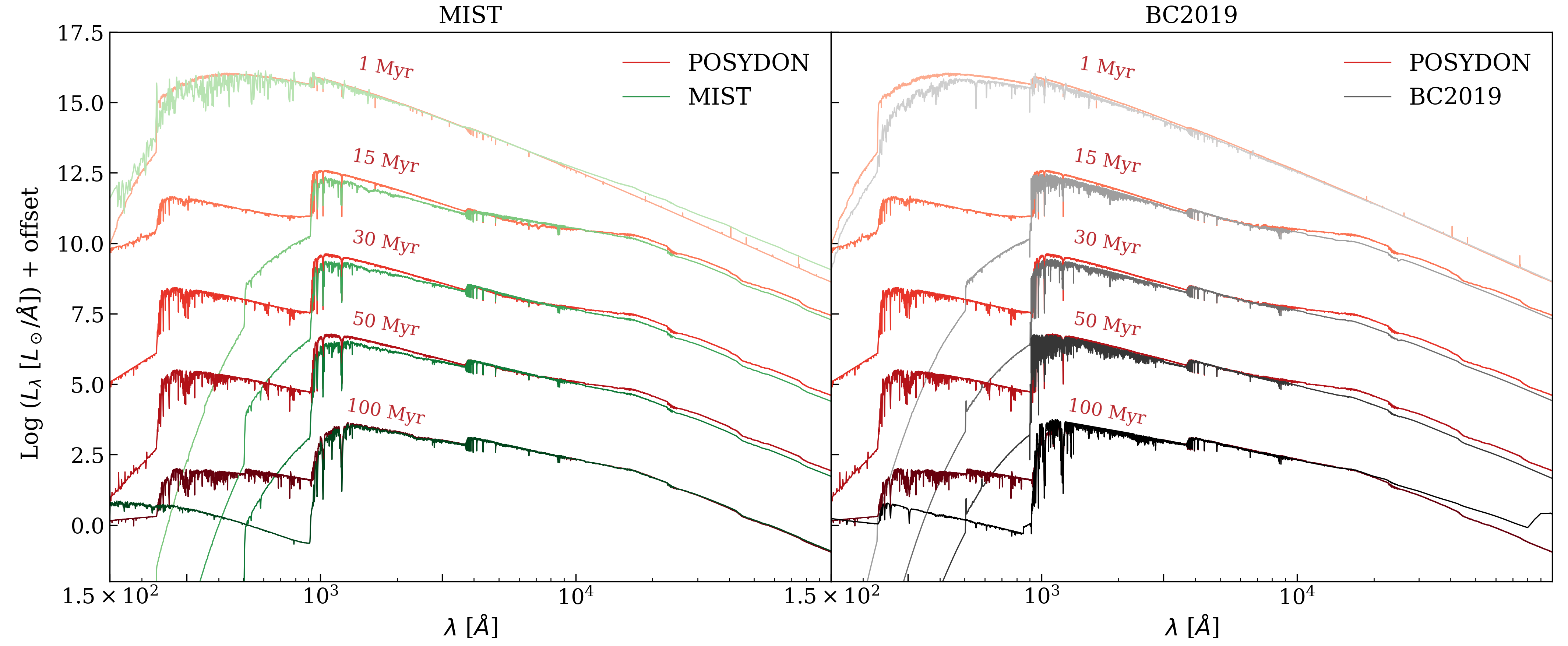}
        \caption{We compare our synthetic stellar population spectra at $Z_{\odot}$ for ages in range of 1 Myr to 126 Myr with single star spectra. The left panel includes spectral models based on the MIST evolutionary tracks, that are colored in green. The right panel has comparisons of the BC2019 models in gray.}
        \label{fig:MIST_PRSC_comp}
\end{figure*}

\begin{figure*}
    \centering
    \includegraphics[width=1\textwidth]{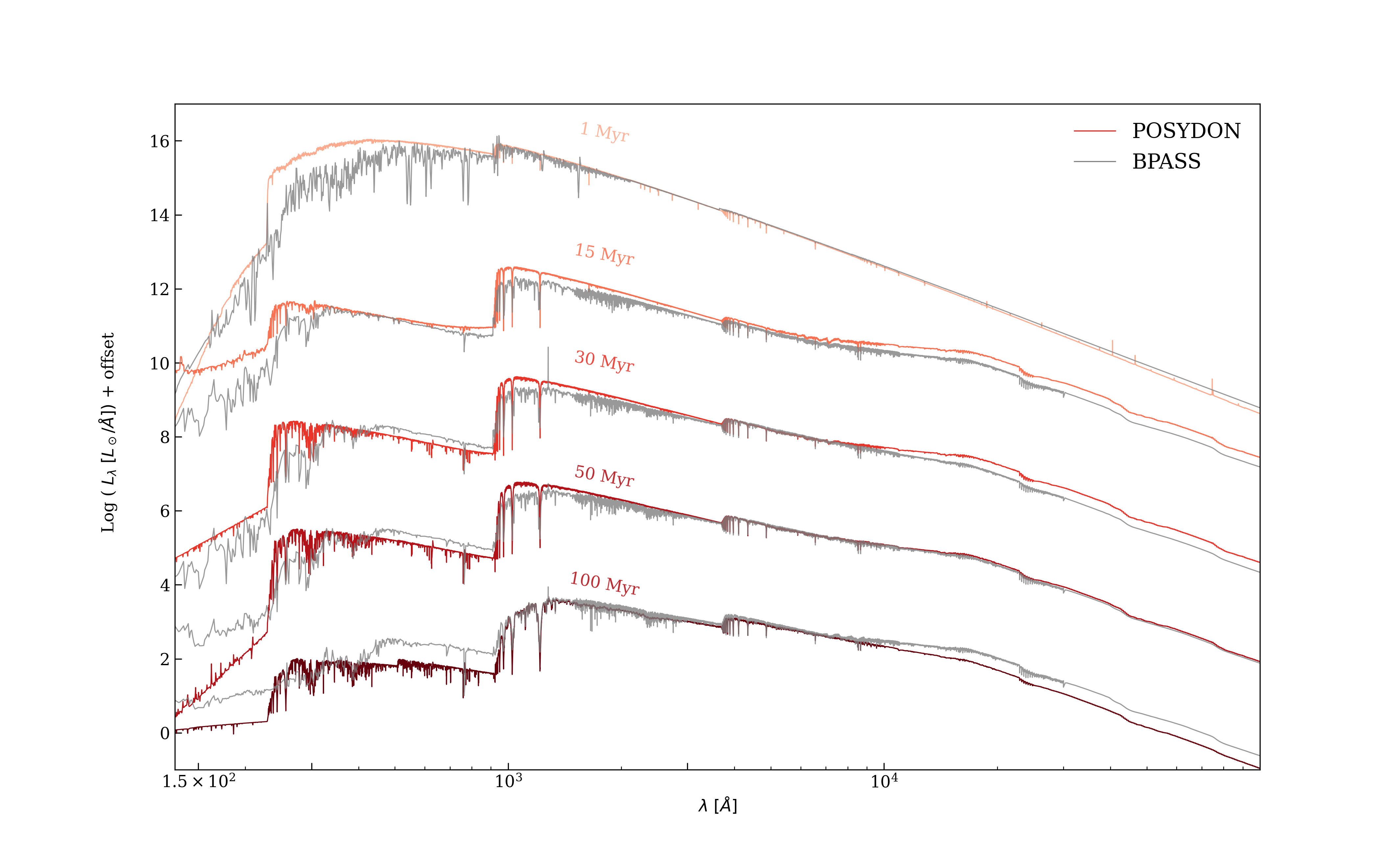}
    \caption{ Comparison of the integrated binary population spectra predicted by our fiducial models and {\tt BPASS} at selected ages. The red spectra correspond our fiducial binary population models, while the grey show the corresponding {\tt BPASS} binary spectra at ages of 1, 15, 20, 50, and 100 Myr. Both spectra represent a population of initial stellar mass of $10^6$. This comparison illustrates how different assumptions about stellar and binary evolution, as well as spectral libraries can produce discrepancies in the UV and LyC of binary populations. }
    \label{fig:BPASS_comp}
\end{figure*}

Our final spectral comparison includes our binary models and those computed with {\tt BPASS} (version 2.2.1) as described in \cite{2018MNRAS.479...75S}. In version 2.2.1, {\tt BPASS} adopts the C3K spectral library for stars below $T_{\rm eff}$ of $25,000~{\rm K}$ and the WM-BASIC spectral library is used for hotter stars. Their WR star treatment includes all stars with $X < 0.4$ and $\log~ T_{\rm eff} \ge 4.45$ \cite[as described in][]{2018MNRAS.479...75S}, and WR stars are assigned as WNL, WNE, or WC subtypes from the PoWR models. 

In Figure \ref{fig:BPASS_comp}, we show the comparison between the {\tt BPASS} spectral models and {\tt POSYDON} models. We choose the {\tt BPASS} spectral models computed with the Salpeter IMF within the mass interval of ($0.1~M_\odot - 100~M_\odot$). The {\tt BPASS} initial binary parameter distributions of mass ratio, period, and binary fraction follow from \cite{2017ApJS..230...15M}. Consequently, the initial parameters of the {\tt BPASS} models should be very close to our own initial parameters.

In Figure~\ref{fig:BPASS_comp} the shape of the overall spectral distributions between the two models match well, with some modest, noteworthy differences, particularly in the UV and IR. First, we see that at wavelengths shorter than 912~\AA, the {\tt BPASS} spectra produce substantial emission, similar in magnitude to our binary spectra. For ages greater than 15 Myr, this emission is clearly a binary evolution effect. However, the shapes of the spectra differ between the two models. We attribute this to the different choices of spectral models applied: the {\tt BPASS} models use WR atmospheres for all H-poor stars above a certain temperature ($log(T_{eff} ) \ge 4.45$), whereas we adopt varying spectral models for atmospheres with optically transparent winds. Self-shielding by winds in WR stars accounted for in the PoWR models leads to a somewhat diminished production of radiation at wavelengths of 250~\AA--450~\AA\  \citep{2018A&A...615A..78G,2022IAUS..366...21S}. Binary stripped star spectral models from \cite{2023ApJ...959..125G} show enhanced emission here, as these stars lack strong winds.

For wavelengths ranging from 450~\AA\ to 912~\AA, the {\tt BPASS} models produce modestly more emission, which we suspect is due to differences in the binary evolution prescriptions--and therefore in the number of stripped stars produced (see discussion below). These minor differences between the two models can influence the emission rates of ionizing species, as well as predictions for emission line ratios such as \ion{He}{2}/${\rm H}{\beta}$.

For UV wavelengths longer than 912~\AA, we again see an excess in the {\tt POSYDON} models at ages less than 100 Myr, which we also suspect is due to the choice of overshooting in {\tt POSYDON}.

Finally, we see differences in the IR ($\geq 1 \mu m$), where in the 15 Myr and 30 Myr models, {\tt POSYDON} shows a comparative excess, but at 100 Myr, a deficit. Since this emission is predominantly generated by giant stars (at this age, specifically thermally pulsating AGB stars), we suspect differences in the physics of AGB pulsations could play an important role \citep[see more in][]{2017PASA...34...58E}. We also note that these stars are short-lived, and depending on the resolution of the underlying stellar grid, stochastic effects can also contribute (see more in Appendix \ref{sec:Appendix C}).

In Figure~\ref{fig:stripped_comp} we present the number of stripped stars from a starburst population as a function of the starburst's age, from both the {\tt POSYDON} and the {\tt BPASS} models. We note that although {\tt BPASS} does not differentiate between optically thin and optically thick H-poor stars they record a population of WNH H-poor stars with luminosities below $\log (L/L_\odot) \le 4.9$. These are the stars which we highlight as `` binary stripped'' in generating the comparison shown in Figure~\ref{fig:stripped_comp}. We find that our models produce approximately half the number of binary stripped stars present in the {\tt BPASS} models. We also include in Figure~\ref{fig:stripped_comp} the binary stripped star predictions from \citet{2019A&A...629A.134G}, that were produced using a Kroupa IMF and a period distribution from \citet{2012Sci...337..444S}. 
Since our adopted initial distributions differ from those of \citet{2019A&A...629A.134G}, we do not expect the predicted number of stripped stars to match, but it can rather serve as a qualitative comparison. We find that \citet{2019A&A...629A.134G} produces a flatter evolution between 40-100 Myr compared to the {\tt POSYDON} and {\tt BPASS} models. Finally, despite the different IMF and period distribution assumptions, the models of \citet{2019A&A...629A.134G} also cannot reproduce the number of stripped stars present in the {\tt BPASS} populations.

While we cannot account for the exact source of differences in the numbers of stripped stars, we point out that stripped stars are a phenomenon produced by binary evolution, either through stable or unstable mass transfer. Therefore different assumptions made about the stability of mass transfer, the mass transfer rate, and the CE treatment can all account for the different outcomes displayed in Figure~\ref{fig:stripped_comp}. These differences also suggest that the growing population of observed stripped stars and their properties have the potential to place invaluable constraints on binary evolution physics.

\begin{figure}
    \centering
    \includegraphics[width=1\linewidth]{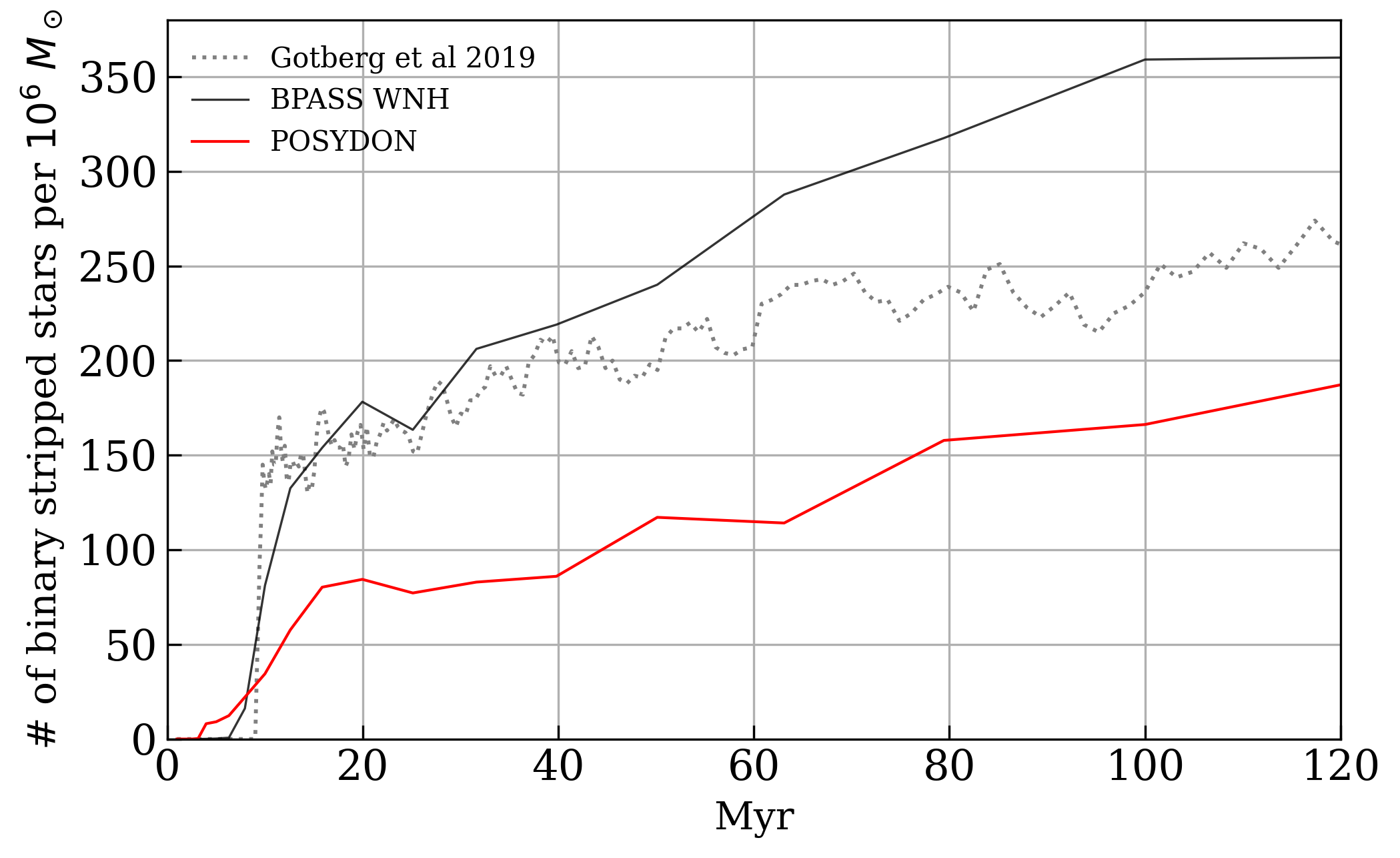}
       \caption{The number of binary stripped stars as a function of time from the {\tt POSYDON} models (Red line) and the {\tt BPASS} models (Black line) as a function of age. For reference we include the predictions of stripped stars from \cite{2019A&A...629A.134G}, albeit they were produced by a different IMF.}
    \label{fig:stripped_comp}
\end{figure}

\begin{figure*}
    \centering
    \includegraphics[width=1\linewidth]{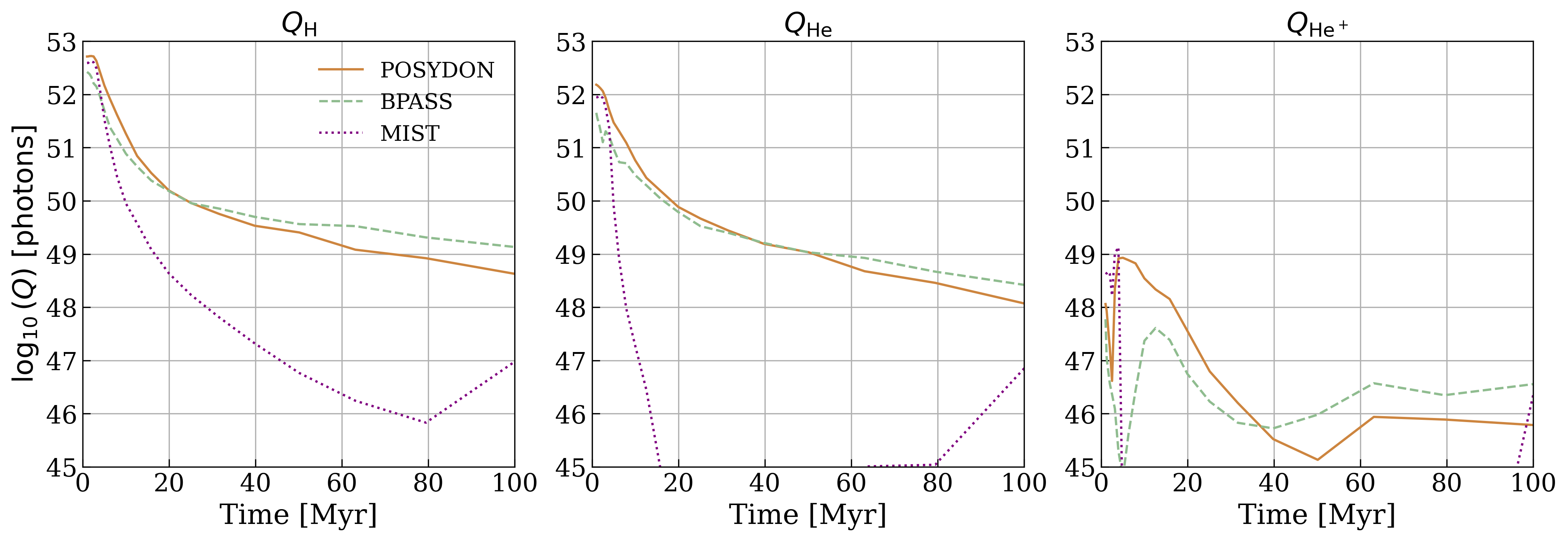}
    \caption{The emission rates of ionizing photons from a starburst as a function of time from three different models: {\tt POSYDON} populations including binaries (solid, orange), {\tt BPASS} populations including binaries (dashed, green), and single-star MIST (dotted, purple). The {\tt POSYDON} and {\tt BPASS} models produce photon emission rates of the same order of magnitude for $Q_{\rm H}$ and $Q_{\rm \text{He}}$, {while they present} the biggest differences in the $Q_{\rm \text{He}^+}$. On the other hand, the MIST spectral models produce significantly fewer ionizing photons compared with the binary models, as binary interactions significantly alter the FUV radiation of stellar populations.}
    \label{fig:ionizing_Comp}
\end{figure*}

As a final comparison, in Figure~\ref{fig:ionizing_Comp} we present the emission rates of key ionizing photons from the {\tt POSYDON} models (solid line), the {\tt BPASS} models (dashed line) and the single-star MIST models (dotted line). The {\tt BPASS} and {\tt POSYDON} ionizing outputs for $Q_{H}$ and $Q_{\text{He}}$ are within the same order of magnitude, and substantial differences appear in some ages ($\sim 80$ Myr). The differences are more pronounced however in the $Q_{\text{He}^+}$ ionizing emission rate. As we illustrate in Figure~\ref{Ionizing radiation}, the $\text{He}^+$ ionizing emission is the most sensitive to binary physics and the underlying population of stripped stars. From Figure \ref{fig:stripped_comp} we see that, although the {\tt BPASS} models exceed the number of H-poor stars present in our populations, at ages younger than 40 Myr the {\tt POSYDON} emission rates are higher likely due to our adopted spectral shapes and some key evolutionary parameters like overshooting. After 40 Myr, the excess of stripped stars along with contributions from WD and PN/post-AGB stars in the {\tt BPASS} models likely drive the reversal in the predicted output of $Q_{\text{He}^+}$ between the two models.

\section{Limitations, Caveats, and Future Work}
\label{sec:limitations}

    While our spectral models provide some improvements over other models in the literature, they are subject to a number of limitations, many of which we plan to address in the future.

     First, our spectral models are constructed from the publicly available \texttt{POSYDON} binary grids that include binary systems where the primary has a minimum mass of $4~ M_\odot$. Since we did not model any binary interaction for binaries with primaries below $4~ M_\odot$, one should take caution when interpreting the output of our populations after all the more massive stars have evolved off the main sequence ($\gtrsim$120 Myr). For systems with less massive primaries, we include stellar binaries with astrophysically motivated distributions of mass and mass ratio but ignore any binary interactions. Depending on the use case, these models may be sufficient and will be no worse than single star spectral models. However, without the inclusion of mass transfer and stellar mergers, our models will be devoid of blue stragglers and stripped stars at these older ages. We therefore suggest that our models will under-predict the production of blue light, especially ionizing radiation, after 120 Myr. 

    Second, since in this work our focus has been to address the output of massive stars, we defer the treatment of WDs and post-AGB stars to future versions of our spectral models. The contribution of these systems, although minor for young populations where the production of blue radiation is dominated by high-mass stars, will start having an effect on our population spectra at the shortest wavelengths ($\lambda \lesssim 912$ \AA) after 80 Myr. We have some stochastic variability in our models with ages $\ge 80 \, \text{Myr}$ due to short-lived phases of stellar evolution that fall just outside of our post-AGB criteria.

    We note that the stellar models in {\tt POSYDON} track the mass loss of AGB stars as they evolve into WDs. The stars can briefly become much hotter than massive OB stars. Observationally these stars are known to be surrounded by the ejected AGB envelope, which can absorb a significant fraction of their ionizing radiation. Evolutionary models of post-AGB stars show that the PN can transition from being optically thick to optically thin to the Lyman continuum \citep{2004A&A...423..995M, 2007A&A...473..467S}, leading to the leakage of ionizing photons. It is therefore important that we not overestimate their LyC contribution. In this work we have removed these systems from our populations by hand, but in future works we hope to incorporate a more physical model properly accounting for local gas reprocessing. We also note that post-AGB stars are so short-lived that stochastic effects must be also taken into account.
    
    When we inspect the population of binary stripped stars, we find that many systems have moderate surface H abundances ($X \approx 0.4$), as mass transfer only partially stripped the star.
    These stars have larger radii than the weak-wind H-poor spectral models from \cite{2023ApJ...959..125G}, and often map to regions outside of their stellar libraries. We currently map to the closest grid point, which typically corresponds to model with an effective gravity ($\log g$) approximately 10\% larger. We therefore expect our models will slightly underestimate their contribution in the LyC. A more complete model would expand the spectral library of stripped stars which requires new radiative transport simulations of stellar envelopes, a task outside the scope of this work.
    
    Another potential limitation of the {\tt POSYDON} populations is the absence of modeling the mass transfer between a post-CE exposed He-core with a MS star. As {\tt POSYDON} lacks a {\tt MESA} grid to evolve such binaries, these systems are halted upon Roche lobe overflow. These systems enter this phase of mass transfer  when the He-star is very close to core collapse, and correspond to about $~19\%$ of the post-CE systems or $~4\%$ of the overall population of high mass interacting binaries. We expect the exclusion of these systems to have little effect on our population spectra. However, we note that in a fraction of these systems after the primary has become a CO, the secondary star could go through its own stripped phase upon evolving off the main sequence. While, again, we expect the overall contribution of these systems to be minor, a more complete model will account for this formation channel.

    Finally, {\tt POSYDON} is missing a key evolutionary channel for the production of chemically homogeneous evolving stars \citep{2007A&A...465L..29C,2023MNRAS.518..860G}, which are spun up to near critical rotation through mass accretion.
    This deficiency exists because even though {\tt POSYDON} incorporates the spin of both stars within a {\tt MESA} grid, after the primary has gone through core collapse, {\tt POSYDON} uses a single star grid with no rotation to model detached systems with a star and an compact object companion. Accretion induced rotation enhances the mixing in the stellar interior, and is suggested to lead to chemically homogeneous evolution. While such systems would produce significant blue radiation, the parameter space where this occurs is expected to be small at low metallicities \citep{2005A&A...443..643Y} and non-existent at Solar metallicity, the focus of this work. Future studies expanding our models to lower metallicities will determine exactly how important this channel is for the production of stellar population spectra.
    

    All stellar population spectra incorporate approximations, and, for the limitations previously described, our model is no exception. In future versions of our binary population spectra, we hope to address several of the above limitations by expanding to lower metallicities \citep[provided by {\tt POSYDON} version 2;][]{2025ApJS..281....3A} and lower masses (and therefore older ages). At the same time we plan on addressing other deficiencies in our model by incorporating the contribution from e.g., white dwarfs and post-AGB stars. 
    Nevertheless, our spectral models described in this work are optimized for many use cases. Our models are applicable for starburst galaxies dominated by recent bursts of star-formation (ages $\lesssim100$~Myr) or young star-clusters, with a particular focus at blue wavelengths. To highlight one possible application, our spectra are particularly relevant for self-consistently modeling the high mass X-ray binary populations and their integrated X-ray emission in the context of the stellar populations in these starburst galaxies \citep[see, e.g.,][]{2024ApJ...960...13G, 2025MNRAS.542.2087B}.
    

    \section{Summary \& Conclusions}
    \label{sec:summary}
    In this work we presented our framework for creating spectral models from binary populations models using \texttt{POSYDON}. We evolved a co-eval population of $10^6 M_\odot$ at solar metallicity, and computed the integrated spectrum over a range of ages spanning from 1 Myr to 15.6 Gyr. We compared our spectral models to a {\tt POSYDON}-only single-star population and with two different single-star spectral models from the literature, as well as with spectral models from the binary spectral code {\tt BPASS}. Below, we summarize our key findings: 

        \begin{itemize}
            \item Our binary models confirm what was already established by other works: that binary interactions play an important role in shaping the SEDs of stellar populations. By comparing our models with several single-star populations, we show that binary populations produce and sustain an excess flux in the UV, especially the FUV, in the first 100 Myr. 
            
            \item The sources that are responsible for the majority of the FUV emission are primaries that were stripped through mass transfer. We find that the majority of these stripped stars in our populations are produced through stable case B mass transfer. 
            
            \item  We find that mergers account for a substantial fraction of the overall UV luminosity, especially at shorter wavelengths. 

            \item We also compare the emission rates of ionizing photons predicted by single and binary star models. We find that the $\text{He}^{+}$ emission rate in particular is the most sensitive to the inclusion of binary evolution, showing the strongest enhancement relative to single star populations.

            \item The method we use to produce our populations differs from other population spectral codes that are calculated from the weighted contribution from different models within a grid. We demonstrate that, depending on the grid resolution, such methods can suffer from stochastic effects. Our method mitigates some of these effects by sampling large numbers of stars using {\tt POSYDON} to interpolate between grid points, and then summing their individual spectra.

            
        \end{itemize}

    The models described in this work are produced from open source code that has been integrated into the {\tt POSYDON} repository. The population spectral models we generate have been packaged for easy integration into well-known spectral fitting codes within the community including {\tt FSPS}, {\tt PROSPECTOR}, {\tt BAGPIPES}, and {\tt LIGHTNING}. However, for users who would like to evolve their populations (for instance for modeling stellar population at ages in between the ones provided in our pre-computed grid) documentation within the {\tt POSYDON} repository provides instruction on producing new data sets.


            





\begin{acknowledgments}
We thank the referee for a careful review of the manuscript and useful comments which greatly improved its quality.
We would like to acknowledge useful discussions with  Miguel Cervi\~{n}o. The {\tt POSYDON} project is supported by the Gordon and Betty Moore Foundation (PI Kalogera, grant awards GBMF8477 and GBMF12341) and a Swiss National Science Foundation (PI Fragos, project numbers PP00P2\_211006 and CRSII5\_213497). E.K.\ and J.J.A.\ acknowledge support for Program number (JWST-AR-04369.001-A) provided through a grant from the STScI under NASA contract NAS5-03127. R.H.D.T.\ acknowledges support from NASA ATP award 80NSSC24K0895 and NSF AAG award 2407636. The authors acknowledge UFIT Research Computing \url{http://www.rc.ufl.edu} for
providing computational resources and support that have contributed to the research results reported in this paper.
This research was supported in part through the computational resources and staff contributions provided for the Quest high-performance computing facility at Northwestern University which is jointly supported by the Office of the Provost, the Office for Research, and Northwestern University Information Technology.
\end{acknowledgments}

%

\vspace{5mm}


\software{ This manuscript made use of the following Python modules: {\tt numpy} \citep{harris2020array}, {\tt matplotlib} \citep{Hunter:2007}, {\tt astropy} \citep{astropy:2013,astropy:2018,astropy:2022}, {\tt scipy} \citep{2020SciPy-NMeth}, {\tt pandas} \citep{reback2020pandas} }



\appendix

\section{Time sampling of {\tt POSYDON} binary populations}
\label{sec:Appendix A}

\begin{figure*}
    \centering
    \includegraphics[width=1\linewidth]{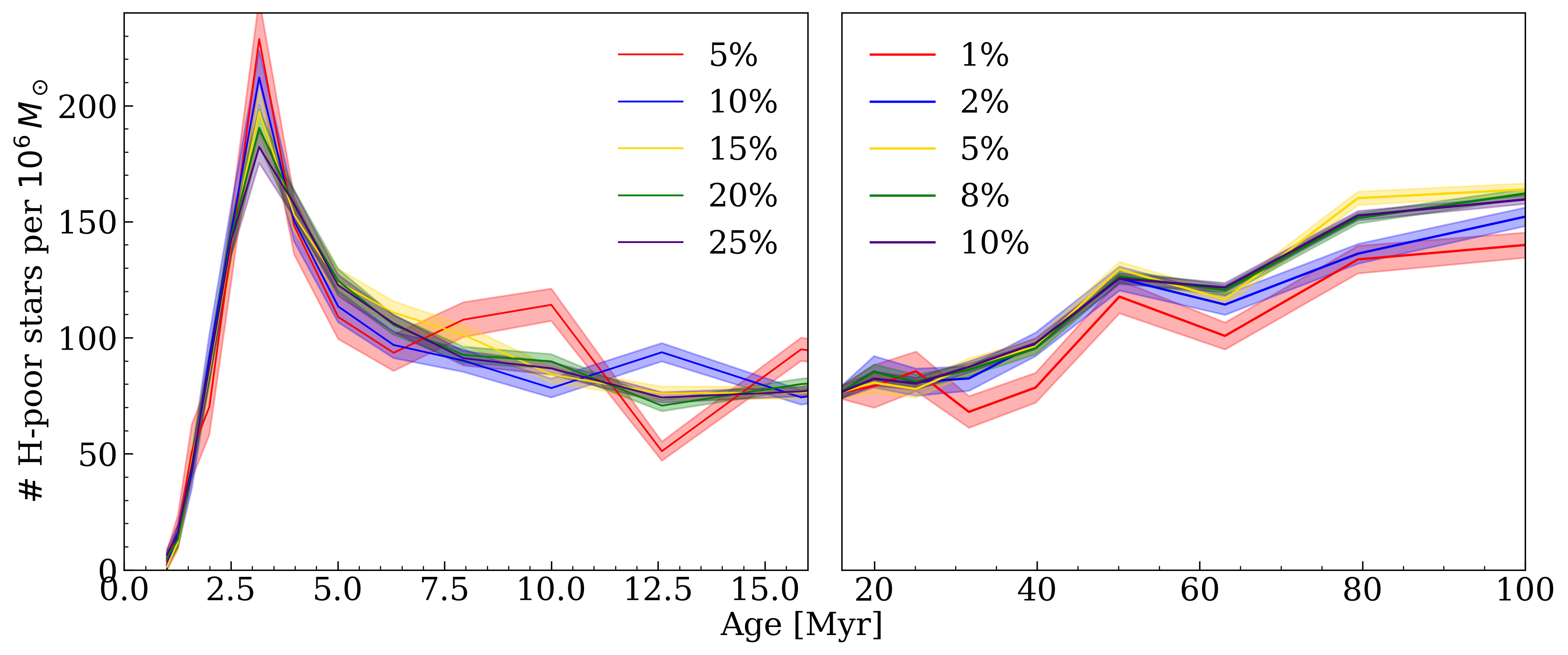}
    \caption{The number of primary H-poor stars for starbursts of $10^6 \, M_\sun$ across a range of ages. We have chosen to separately explore the convergence in two different regimes: one for ages less than 16 Myr (left panel) and one for ages older than 16 Myr (right panel). For each age we assume a time binning equal to a percentage of that age. Increasing the bin widths leads to convergence in our predictions of stripped stars. However for bins too wide, real features of the population are artificially smoothed out. We optimize these competing effects by choosing a bin width of $\pm15\%$ for ages less than 16 Myr and $\pm8\%$ for ages above 16 Myr.
    }
    \label{fig:convergence_N_stripped_stars}
\end{figure*}

\begin{figure*}
    \centering
    \includegraphics[width=\linewidth]{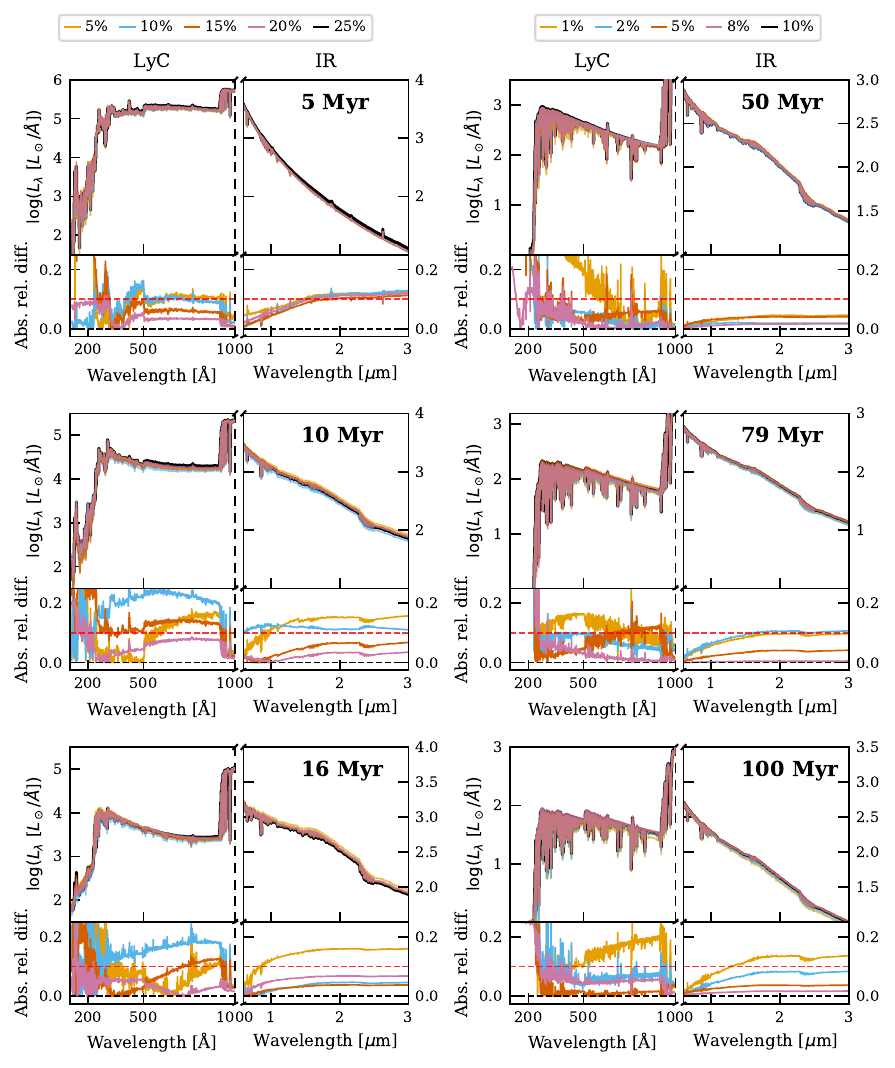}
    \caption{Comparison of starburst spectra computed using bin widths to illustrate the effect of time binning on population spectra. The left column shows populations constructed with bin widths ranging from $ \pm10\%$ to $25\%$, while the right column displays narrower bin widths from $\pm1\%$ to $10\%$. We isolate the two wavelength regimes exhibiting the largest spectral variations: the Lyman continuum (LyC) and the infrared (IR). The lower panels show the residuals of the flux from each bin width setting, compared with that of the largest bin width which is used as the reference. The red dashed line denotes a 10\% residual value.
    The residuals decrease with increasing bin width, demonstrating the convergence of the population spectra with respect to the adopted time binning.}
    \label{spec_binning}
\end{figure*}

{\tt POSYDON} generates populations from the pre-calculated {\tt MESA} grids that are regularly spaced in three initial parameters: primary mass $m_1$, mass ratio $q$, and initial period $\log P$. As mentioned in Section~\ref{sec:Methodology} we utilize a classification and interpolation method to predict the final outcome of binary evolution within these grids. While interpolation of the final values in {\tt MESA} binary grids is possible, time-dependent interpolation---i.e, track interpolation within our binary grids---is currently in the process of being implemented in the code \citep{2025ApJ...984..154S}.

Evolutionary stages within {\tt POSYDON} that are integrated through the detached step \citep[see][for more details]{2023ApJS..264...45F} are calculated by interpolating single star models using equivalent evolutionary points. In these cases, the binary's evolution is properly interpolated. However, to obtain stellar and binary parameters of a system at a specific age while its evolution is still within the binary grids, we resort to nearest-neighbor (NN) matching; we are able to interpolate in time but not mass. Although our grid resolution is sufficiently fine to accurately account for the wide range of possible binary interactions, the necessity of having grids with finite spacing introduces limitations in accurately predicting short-lived stellar phases.


This limitation becomes especially relevant for post-MS stellar evolution, when the donor is an AGB and/or stripped star, phases with significant impact for the spectral outcome of a binary population. Stars that end up losing part or all of their envelopes only spend $\simeq$10\% of their MS lifetimes in the stripped star phase.

As shown in Figure \ref{stripped}, Case B mass transfer produces the majority of stripped stars. Since our {\tt MESA} grids are regularly spaced, each model within a given mass slice will begin to exit their MS and cross the Hertzsprung gap at approximately the same time and will spend a similar amount of time in their stripped star phase (about 10\% of the star's lifetime). Since these phases often occur during a {\tt MESA} grid, the stellar properties at a given time are extracted by the NN method. If the time that is needed for two consecutive mass slices to evolve past their MS is longer than the duration of the stripped star phase, then there will be a period in time where NN interpolation will artificially not find any stripped stars. This issue is not unique to our approach, but it is a problem that all regularly spaced grids inherit.

As we demonstrate in Section~\ref{sec:ionizing_radiation}, the total number of stripped stars will determine the hardness of the ionizing radiation of binary populations; therefore, it is crucial to minimize the potential of numerical artifacts arising from our grids. To address this issue, the Version 2 {\tt POSYDON} HMS-HMS grid has an increased resolution in the mass range spanning $ 4 - 14 \, M_\odot$. 

To calculate the spectrum produced by a stellar population at a particular age, we apply a novel approach; whereas most spectral synthesis codes calculate their population spectra by extracting the stellar properties from a grid of stellar evolutionary tracks at the target age, we randomly draw large numbers of stars, each simulated to a random age from within a uniform ``bin'' around the age we seek to model.
This approach has the practical effect of softening any numerical effects resulting from our choice of adopting a regularly spaced grid in mass.

We determine the appropriate age bins width for our populations by performing a convergence test. We use the number of H-poor primary stars (both WR and stripped stars) produced as a function of time in coeval populations with varying age bin widths representing individual bursts. Since our resolution is increased for stars with masses below $14~ M_\odot$, we assumed two different bin sizes: one for bursts younger than 16 Myr, corresponding to the lifetime of a $14~ M_\odot$ single star model and another for bursts that are older than that. For both age regimes, we selected the smallest possible bin width that minimizes stochastic variations. The width of the bins is calculated as a percentage above and below a given age.

We selected a subset of the starburst ages chosen to evolve our populations between 1 Myr to 100 Myr, for which the effects of binary interactions are not impacted by the grids lower mass limit of the binary grids. For each of the 21 ages equally spaced in log space, we vary the assumed bin width and record the number of H-poor stars found in each population. Figure \ref{fig:convergence_N_stripped_stars} shows the number of H-poor stars in the co-eval populations of different bin widths as a percentage of that age.

For ages younger than 16 Myr, the number of H-poor stars begins to converge for age bin of width 15\% and larger, whereas for ages older than 16 Myr, convergence is achieved at smaller bin widths of about 8\%. Smaller bin widths (e.g., see the red lines in both panels of Figure \ref{fig:convergence_N_stripped_stars}) show unphysical variations in the number of stripped stars over time. We find that a bin size of 15\% for populations whose starburst age is less than 16 Myr and a bin width of 8\% is sufficient to capture stars during their H-poor phase, as increasing the bin width does not change the resulting number of H-poor stars found. Based on this, we adopt a bin width of 15\% for all the populations younger than 16 Myr and 8\% for older populations to construct our co-eval populations used to generate our spectral models.

In addition to the overall number of H-poor stars, the second convergence test we perform examines the effects of different bin widths on the emitted flux from our starburst models. We compute spectra from populations constructed with a range of bin widths and assess the convergence of the emergent flux with increasing bin size. Figure \ref{spec_binning} presents our comparisons in two different wavelength regimes where the largest variations are observed; the LyC ($\lambda < 1000$ \AA) and IR ($1-3~ \mu m $). The variability in the LyC continuum is driven by stripped stars, and in the IR by post-MS stars, both of which are relatively short-lived stellar phases. In contrast, the optical regime exhibits minimal variations, as the contributions from both younger and older MS stars are averaged within that bin.

The left column of Figure \ref{spec_binning} shows populations younger than 16 Myr, for which the adopted bin widths range from 5-25\%, while the right column includes older populations with narrower bin widths of 1-10\%. To provide a better qualitative view of the convergence, we plot the residuals with respect to the population constructed using the biggest bin width. We find that the residuals decrease with increasing bin width, indicating that the spectral variations are progressively smoothed out. With the exception of the smallest bin width populations, and for most wavelengths, the residuals remain below the red, dashed line representing a 10\% relative difference. Consequently, our adopted bin width choices of 8\% and 15\% produce spectra that are sufficiently close enough to those of larger bin widths. We note, however that significant differences remain for wavelengths smaller 200 \AA, indicating that emitted \ion{He}{2} ionizing photons are subject to larger uncertainties.

\section{Treatment of H-poor stars}
\label{sec:H-poor treatment}
The radiative feedback of H-poor stars is important for stellar populations, especially when it comes to their hard, ionizing radiation output. In this work, we consider two different classes of H-deficient stars: weak-wind H-poor stars and WR stars. In our spectra, these are modeled separately and are assigned to different spectral libraries. We will briefly define the distinctions between the two and how our library selection process decides which library is more suitable for each star.

Wolf-Rayet (WR) stars are massive H-poor stars that exhibit strong winds. 
These radiatively driven winds produce the characteristic emission-line spectra that classify them as WR stars \citep[see][for a review]{2007ARA&A..45..177C}. It has been shown from observations \citep{2020A&A...634A..79S} that there is a metallicity-dependent luminosity cut-off below which no WR stars are found. Stars below this threshold can still become H-poor, but they have optically transparent winds and they are not expected to produce WR-type spectra \citep{2020MNRAS.499..873S,2022arXiv221105424S}. Such stars require a different treatment \citep{2019A&A...629A.134G} and are classified as weak-wind H-poor stars in our populations.


At solar metallicity single stars more massive than $\sim20 M_\odot $ \citep{2019A&A...627A.151S} can achieve the necessary wind mass-loss rates and luminosities typically required for a star to become a WR. Weak-wind H-poor stars often reside in the intermediate mass regime, between WR stars and hot sub-dwarfs, and they are the direct products of binary interactions \citep{1992ApJ...391..246P}. These binary-stripped stars have optically thin atmospheres, weak winds, and they exhibit various amounts of hydrogen in their spectra \citep{2023ApJ...959..125G,2023A&A...674L..12R}. They are thought to spend most of their evolution as hot stars, emitting the majority of their radiation in the ionizing regime \citep{2018A&A...615A..78G}. The primary mechanism which these stars lose their hydrogen envelope is through mass stripping from binary interactions leaving them depleted of H, with the hot, compact He-core exposed \citep{1969A&A.....3...83K,2017A&A...608A..11G}.

In some cases, however, mass transfer has not removed the entirety of the envelope of the stripped star, resulting in stars with larger radii and cooler temperatures \citep[so called ``puffed-up binary stripped stars'';][]{2022A&A...662A..56K}. Both stripped and partially stripped, puffed-up stars appear in our populations, but we do not make a spectral distinction between the two and assign them both to the same weak-wind H-poor star library. Since our weak-wind H-poor spectral library accounts for a varying range of $X$ (see Table~\ref{tab:spectral_grids}), our model properly accounts for this variation. However, for puffed-up, stripped stars with effective temperatures below the lower limit of the weak-wind H-poor library (30,000 K), we opt to calculate their spectra using O or B-type stellar libraries. 

The primary, physical distinction we consider separating WR stars from weak-wind H-poor stars is whether their stellar properties imply the presence of optically thin or thick winds. To determine this, we follow the approach of \cite{2022A&A...661A..60A} and calculate the optical thickness $\tau(R)$ of the winds as defined by \cite{1997ApJ...477..792D}: 

\begin{equation}
    \tau (R) = \frac{- \kappa \dot{M}}{4 \pi R (v_\infty - v_0)} \ln \left( \frac{v_\infty}{v_0}\right),
    \label{tau}
\end{equation}
where $\kappa$ is the electron scattering opacity, $\dot{M}$ is the mass loss rate, $R$ is the radius of the star, $v_\infty$ is the terminal velocity, and $v_0$ is the sound speed. 

We calculate the terminal velocity for a star of mass $M$ from \cite{2002A&A...387..244G}: 
\begin{equation}
    v_\infty = 1.3 \sqrt{\frac{2GM}{R} (1 - \Gamma)}
\end{equation}
and $\Gamma$ is the Eddington factor given by: 
\begin{equation}
    \Gamma = \frac{\kappa L}{4 \pi c G M},
\end{equation}
where $G$ is the gravitational constant.

These equations give us an estimate for the optical thickness of the winds, but the quantity being calculated here does not necessarily correspond to the observed optical thickness of these atmospheres. We use different values of optical thickness for stars that are H-free and those that have varying degrees of hydrogen in their atmosphere. \citet{2022A&A...667A..58P} found that adopting $\tau \ge 0.5$ to distinguish between H-poor WN and weak-wind stars yielded good agreement between their model predictions and observations in the LMC. From our single-star populations we see that stars with mass loss rates and luminosity values characteristic of WR stars have optical depths as low as $\tau = 0.3$. In addition, \citet{2023A&A...672A.198S} find that H core-burning massive WR stars can be formed from Case A mass transfer that have values of $\tau \ge 0.1$. 

    \begin{figure*}
    \centering
    \includegraphics[width=0.9\linewidth]{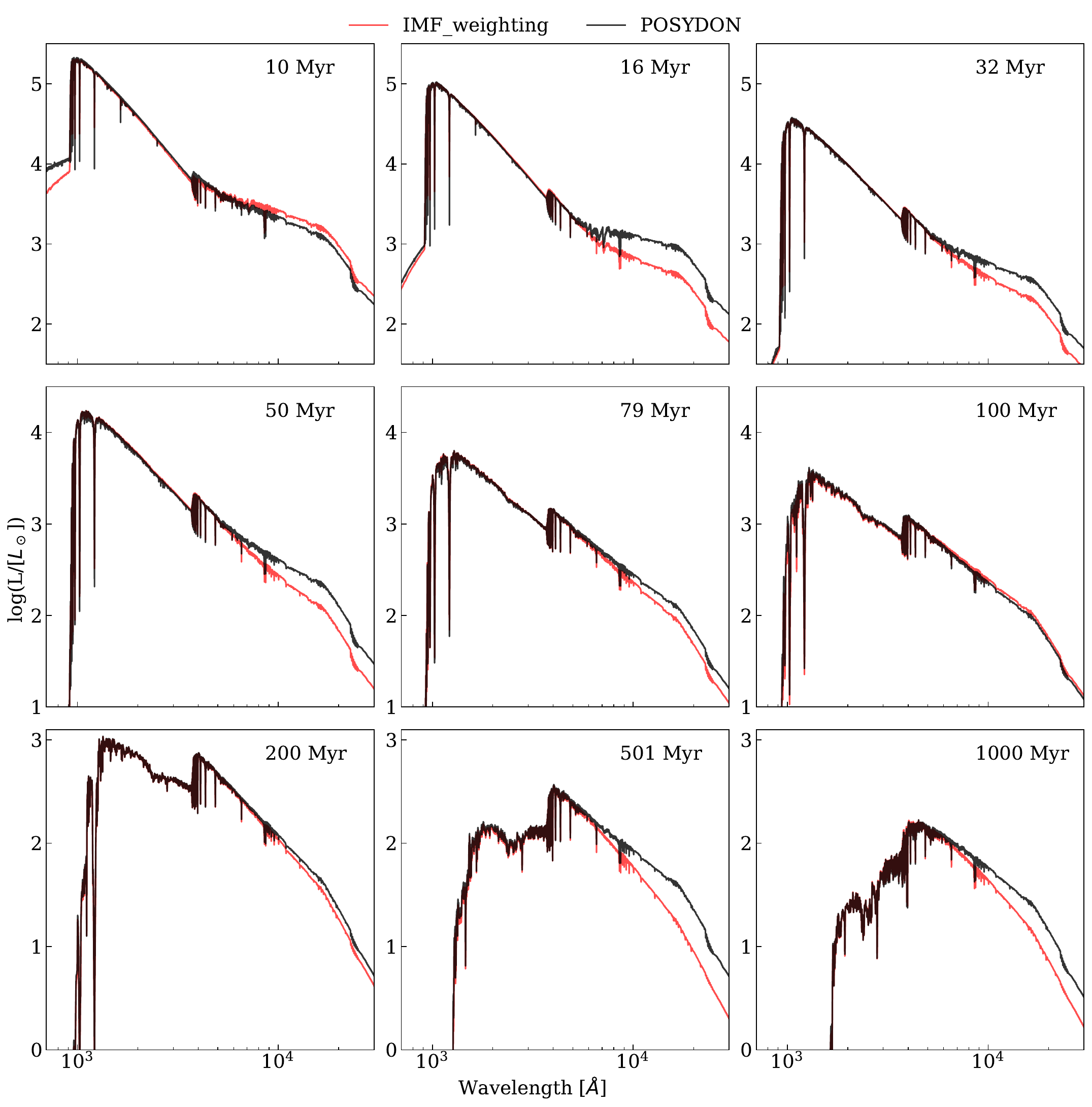}
    \caption{We compare spectral models of several ages made with two different methods. The black lines show the spectra of $10^6 M_\odot$ single star population generated by a combination of stochastic sampling and time-dependent interpolation for the stellar properties. The lines in red are the spectral models for populations of the same stellar mass, that were constructed by weighting the individual spectra of every star in our single-star grids by the IMF. The differences between the two models are in the IR wavelengths, where the spectra produced by IMF produce systematically less flux.}
    \label{weight_comp}
\end{figure*}

Based on these results, we distinguish between hydrogen-poor stars that are still burning hydrogen in their cores (ms-WR) and those that have evolved beyond the main sequence (WRc). The first criterion in our selection process is the presence of H. If a H-poor star retains any amounts of H ($X  > 10^{-2}$) and is still undergoing hydrogen burning, it is assigned to a WNL atmosphere provided that $\tau \geq 0.1$. For all other stars that are found in later stages of their evolution, we select a WNL atmosphere if $\tau \geq 0.3$, otherwise they are assigned to the stripped-star library. 

For hydrogen-free stars, we require an optical depth of $\tau \ge 1$ to assign them to either a WNE or a WC atmosphere depending on their He surface abundance, with WC atmospheres being selected if their helium mass fraction $Y < 0.7 $. The rest of the hydrogen free stars that meet this criterion of the wind optical depth are matched to a stripped-star spectrum. 

The $\tau$ boundary criteria that we selected based on the surface compositions and evolutionary states of H-poor stars, are expected to primarily affect the spectral classification of H-poor stars in populations with ages between 10 and 15 Myr, as these stars have $\tau$ values close to the boundaries. As has been shown in \cite{2022arXiv221105424S} and \citet[Appendix D.2]{2019A&A...629A.134G}, the $\text{He}^+$ photons are the most sensitive to wind assumptions made between different spectral models. We therefore expect the $\text{He}^+$ ionizing radiation predicted for our populations within that age range to be the most sensitive to our adopted optical depth criteria.

\section{IMF weighting vs interpolating in the grids}
\label{sec:Appendix C}

    Historically, population spectra are generated by weighting each model from initial distributions for the stellar and binary parameters, such as the IMF \citep[see discussion in][]{2013ARA&A..51..393C}. In this work, we take a different approach, combining stochastic sampling along with grid interpolation. This choice allows us to evolve stars and binaries with arbitrary initial parameters, though it comes with an increased computational cost.  

    In this section, we compare the spectral outcomes of these two methods, focusing only on populations of single stars. As highlighted in the case of stripped He-stars in Appendix \ref{sec:Appendix A}, short-lived phases of stellar evolution require a sufficiently dense grid in initial stellar mass to resolve. As a reminder, {\tt POSYDON} interpolates single star evolution in both mass and age using equivalent evolutionary points \citep{2016ApJS..222....8D}. Utilizing interpolation not only avoids the need for denser grids by adding more isochrones but it can also reproduce more accurately the physical properties of stars during their post-MS evolution. The major drawback of this method is that it requires simulating large number of stars in POSYDON $\sim \mathcal{O}(10^6)$ and calculating the spectra of every individual star, rather than only computing spectra of each model in the grids which are of the order of $\sim \mathcal{O}(10^3)$ for the single star grid.

    For our comparison, we create an additional set of spectra for each age of coeval populations using IMF weighting. We evolve the 340 single stars in the {\tt POSYDON} grids spanning the mass interval [$0.1-150~ M_\odot$] for the same burst ages, weighting each star's flux contribution according to the Salpeter IMF, normalized to a total stellar mass of $10^6 M_\odot$. In Figure~\ref{weight_comp} we show the spectra produced by the two methods for several ages. For all ages the optical emission is consistent between the two models. However substantial discrepancies emerge in the IR regime which is dominated by post-MS stars. The spectra produced by IMF weighting underproduce the amount of IR radiation for most ages, with the 16 Myr, 50 Myr, 501 Myr, and 1 Gyr models showing the largest deviations. 

    By using interpolation along with stochastic sampling we are able to sufficiently populate the short-lived post-MS evolution at every age. This is particularly important for phases such as the AGB, where a small difference in initial mass between models can result in different effective temperatures and luminosities for the same age. On the other hand, by using IMF weighting of the discrete single star grids only a handful of stars are going to reside in that part of the HR diagram for a particular age. The spectra of these models are going to be disproportionately weighted, causing these individual stars to imprint artificial features on the integrated spectra.


\FloatBarrier

\bibliography{refrences}
\bibliographystyle{aasjournal}



\end{document}